\documentclass[twocolumn,secnumarabic,amsmath,amssymb, nobibnotes, aps, prd]{revtex4-1}

\def\Msun{$M_{\odot}$}

\usepackage{graphicx}
\usepackage{subcaption}
\usepackage{mathrsfs}

\usepackage{xcolor}

\begin{document}

\def\gsimeq{\,\,\raise0.14em\hbox{$>$}\kern-0.76em\lower0.28em\hbox {$\sim$}\,\,}
\def\lsimeq{\,\,\raise0.14em\hbox{$<$}\kern-0.76em\lower0.28em\hbox{$\sim$}\,\,}

\title{Fission fragment distributions and their impact on the r-process nucleosynthesis in neutron star mergers}
\author{J.-F. Lema\^itre}
\email[]{jean-francois.lemaitre@ulb.ac.be}
\affiliation{Institut d'Astronomie et d'Astrophysique, Universit\'e Libre de Bruxelles, Campus de la Plaine CP 226, BE-1050 Brussels, Belgium}
\affiliation{CEA, DAM, DIF, F-91297 Arpajon, France}
\author{S. Goriely}		  
\affiliation{Institut d'Astronomie et d'Astrophysique, Universit\'e Libre de Bruxelles, Campus de la Plaine CP 226, BE-1050 Brussels, Belgium}
\author{A. Bauswein}
\affiliation{GSI Helmholtzzentrum f\"ur Schwerionenforschung, Planckstra{\ss}e~1, 64291 Darmstadt, Germany}
\affiliation{Helmholtz Forschungsakademie Hessen f\"ur FAIR (HFHF), GSI Helmholtzzentrum f\"ur Schwerionenforschung, Campus Darmstadt, Germany}
\author{H.-T. Janka}
\affiliation{Max-Planck-Institut f\"ur Astrophysik, Postfach 1317, D-85741 Garching, Germany}

\begin{abstract}
Neutron star (NS) merger ejecta offer a viable site for the production of heavy r-process elements with nuclear mass numbers $A \gtrsim 140$. 
The crucial role of fission recycling is responsible for the robustness of this site against many astrophysical uncertainties. 
Here, we introduce new improvements to our scission-point model, called SPY, to derive the fission fragment distribution for all neutron-rich fissioning nuclei of relevance in  r-process calculations.
These improvements include a phenomenological modification of the scission distance and a smoothing procedure of the distribution. 
Such corrections lead to a much better agreement with experimental fission yields. 
Those yields are also used to estimate the number of neutrons emitted by the excited fragments on the basis of different neutron evaporation models. 
Our new fission yields are extensively compared to those predicted by the so-called GEF model.

The impact of fission on the r-process nucleosynthesis in binary neutron mergers is also re-analyzed. 
Two scenarios are considered, the first one with low initial electron fraction is subject to intense fission recycling, in contrast to the second one which includes weak interactions on nucleons. 
The various regions of the nuclear chart responsible for fission recycling during the neutron irradiation as well as after freeze-out are discussed. 
The contribution fission processes may have to the final abundance distribution is also studied in detail in the light of newly defined quantitative indicators describing the fission recycling, the fission seeds and the fission progenitors. 
In particular, those allow us to estimate the contribution of fission to the final abundance distribution stemming from specific heavy nuclei. Calculations obtained with SPY and GEF fission fragment distributions are compared for both r-process scenarios.
\end{abstract}

\maketitle

\section{Introduction}

The r-process, or the rapid neutron-capture process, of stellar nucleosynthesis is invoked to explain the  production of the stable  (and some long-lived radioactive) neutron-rich nuclides heavier than iron that are observed in stars of various metallicities, as well as in the solar system (for a review, see Ref. \cite{Arnould07,Arnould20,Cowan20}). 
In recent years, nuclear astrophysicists have developed more and more sophisticated r-process models, trying to explain the solar system composition in a satisfactory way by adding new astrophysical or nuclear physics ingredients. 
The r-process remains the most complex nucleosynthetic process to model from the astrophysics as well as nuclear-physics points of view. 
Progress in the modeling of type-II supernovae and $\gamma$-ray bursts has raised a lot of excitement about the so-called neutrino-driven wind environment. 
However, until now a successful r-process cannot be obtained {\it ab initio} without tuning the relevant parameters (neutron excess, entropy, expansion timescale) in a way that is not supported by the most sophisticated existing models \cite{Wanajo11,Janka12}. 
Although these scenarios remain promising, especially in view of their potential to contribute to the galactic enrichment significantly, they remain affected by large uncertainties associated mainly with the still incompletely understood mechanism responsible for the  supernova explosion and the persistent difficulties to obtain suitable r-process conditions in self-consistent dynamical explosion and neutron-star (NS) cooling models \cite{Janka12,Hudepohl10,Roberts2010,Fischer2010}. 
In particular, a subclass of core-collapse supernovae, the so-called collapsars corresponding to the fate of rapidly rotating and highly magnetized massive stars and generally considered to be at the origin of observed long gamma-ray bursts, could be a promising r-process site \cite{Siegel19b}. 
The production of r-nuclides in these events may be associated with jets predicted to accompany the explosion, or with the accretion disk forming around a newly born central BH \cite{Winteler2012}.

Since early 2000, special attention has been paid to NS mergers  as r-process sites following the confirmation by hydrodynamic simulations that a non-negligible amount of matter could be ejected from the system.
Newtonian \cite{Freiburghaus1999,Ruffert01,Janka99, Korobkin12,Roberts11}, conformally flat general relativistic \cite{Oechslin07,Goriely11b,Just15}, as well as fully relativistic \cite{Kyutoku13,Hotokezaka13a, Bauswein14,Wanajo14, Foucart14} hydrodynamical simulations of NS-NS and NS-black hole (BH) mergers with microphysical equations of state have demonstrated that typically some $10^{-3}\,M_\odot$ up to more than 0.1\,$M_\odot$ can become gravitationally unbound on roughly dynamical timescales due to shock acceleration and tidal stripping. 
Also the relic object (a hot, transiently stable hypermassive NS \cite{Baumgarte00} followed by a stable supermassive NS, or a BH-torus system), can lose mass through outflows driven by a variety of mechanisms  \cite{Ruffert99, Dessart09, Fernandez13, Siegel14, Metzger14, Perego14,Just15}.

Simulations of growing sophistication have confirmed that the ejecta from NS mergers are viable strong r-process sites up to the third abundance peak and the actinides \cite{Just15,Goriely11b,Wanajo14,Roberts11,Bauswein13}. 
The r-nuclide enrichment is predicted to originate from both the dynamical (prompt) material expelled during the NS-NS or NS-BH merger phase and from the outflows generated during the post-merger remnant evolution of the relic BH-torus system. 
The resulting abundance distributions are found to reproduce very well the solar system distribution, as well as various elemental distributions observed in low-metallicity stars \cite{Cowan20}. 
During the dynamical phase of the merging scenario, the number of free neutrons per seed nucleus can reach so high values (typically few hundreds) that heavy fissioning nuclei are produced.
Fissioning results in a robust reproduction of the solar-system-like abundance pattern of the rare-earth elements, as observed in metal-poor stars \cite{Arnould07,Goriely11b,Goriely2015,Cowan20}. 
This supports the possible production of these elements by fission recycling in NS merger ejecta. 
In addition, the ejected mass of r-process material, combined with the predicted astrophysical event rate (around 10\,My$^{-1}$ in the Milky Way~\cite{Dominik12}) can account for the majority of r-material in our Galaxy, \textit{e.g.}  \cite{Qian2000,Metzger2010,Goriely11b,Bauswein14}. 
A further piece of evidence that NS mergers are r-nuclide producers indeed comes from the very important 2017 gravitational-wave and electromagnetic observation of the kilonova GW170817 7,\cite{Abbott17,Abbott17b,Cowperthwaite17,Nicholl17,Chornock17,Waxman18,Watson2019}. 

In this specific NS merger scenario, the neutron richness was found to be so high that heavy fissioning nuclei can be produced. 
For this reason, in this astrophysical site, fission plays a fundamental role, more particularly by {\it i)} recycling the matter during the neutron irradiation (or if not, by allowing the possible production of super-heavy long-lived nuclei, if any), {\it ii)} shaping the r-abundance distribution in the $110 \le A \le 170$ mass region at the end of the neutron irradiation, {\it iii)} defining the residual production of some specific heavy stable nuclei, more specifically Pb and Bi, but also the long-lived cosmochronometers Th and U, and {\it iv)} heating the environment through the energy released. More details can be found for instance in Refs.~\cite{Korneev2011,Korobkin12,Goriely13c,Goriely2015,Eichler2015,Mendoza-Temis2015,Giuliani2018,Mumpower2018,Zhu2018,Holmbeck2018,Wu2019,Vassh2019}.

Despite the recent success of nucleosynthesis studies for NS mergers, the details of r-processing in these events is still affected by a variety of uncertainties, both from the nuclear physics and astrophysics point of view. 
For this reason, we present here a new effort to further improve the description of fission fragment distributions (FFD) obtained on the basis of the so-called Scission Point Yield (SPY) model \cite{lemaitre2019}. 
Sec.~\ref{sec:ffd} is devoted to the description of such improvements. 
A phenomenological correction to the scission distance and a new smoothing procedure for deriving the fission yields are presented in Sec.~\ref{sec:spypres}. 
Since the FFD of relevance for the r-process application correspond to the post-neutron-emission one deduced from pre-neutron FFDs, our procedure for estimating the neutron evaporation is detailed in Sec.~\ref{sec:Nevap}.
As r-process is impacted by neutron captures, sensitivity studies are performed in Sec.~\ref{sec:sensi_evapo} to demonstrate that the neutron emission depends mainly on the primary fission yields.
SPY pre-neutron FFDs are compared with experimental data and the well-known GEF (``GEneral description of Fission observables'') calculations \cite{SCHMIDT2016}  in Sec.~\ref{sec:sys}.

In addition, it has been shown that weak interactions may strongly affect the composition of the dynamical ejecta and thus the efficiency of the r-process 
\cite{Wanajo14,Sekiguchi2015,Goriely15a,Foucart2016,Lehner2016,Radice2016,Bovard2017, Martin2018,Goriely19b,Ardevol19}.
In this case, the initial neutron richness of the ejecta can be significantly reduced and consequently the role of fission, albeit in basically all calculations  a significant fraction of neutron-rich ejecta is present. 
In Sec.~\ref{sec:nucleo}, we re-evaluate the role of fission in NS merger in two distinct scenarios, namely the traditional one neglecting weak interactions on nucleons and an updated version of such a simulation where neutrino interactions are included phenomenologically \cite{Goriely15a,Goriely19b} to reproduce qualitatively the results found in Ref. \cite{Ardevol19}. 
In Sec.\ref{sec:finalabund}, we will present the final abundance distributions obtained within both scenarios as well as the role of the fission recycling and link them to fundamental initial properties of ejected material, namely the  initial neutron mass fraction, electron fraction and entropy.
The relevance of the fission recycling and its contribution to the final r-abundances is thoroughly detailed in Secs.~\ref{sec:FR}--\ref{sec:FP} for both scenarios. Finally, conclusions are drawn in Sec.~\ref{sec:con}. 

\section{Fission fragments distributions}
\label{sec:ffd}
\subsection{SPY model}
\label{sec:spypres}

The SPY model is a static and statistical scission point model \cite{lemaitre2015,lemaitre2019} where a thermodynamic equilibrium at scission is assumed, hence the evolution between the saddle and the scission points is neglected. 
The model is based on two pillars, namely the absolute available energy balance at the scission configurations and the statistical description of the available phase space.
The available energy balance is performed for all energetically possible fragmentations of a fissioning system at scission as a function of the deformation of both fragments. 
The available energy is defined as the difference between the potential energy of the fissioning system at scission and the energy of the excited compound nucleus where both nascent fragments are supposed to be at rest. 
 The potential energy of the fissioning system at scission is obtained as the sum of the individual binding energies of the two fragments and the interaction energy between the fragments composed of the Coulomb repulsion and the nuclear attraction. 
The system at scission is treated as a microcanonical ensemble where all available states are equiprobable. 
In this framework, the number of available states of a given fragmentation is the product of the state densities of the two isolated fragments. 
The yield of a fragmentation is the number of available states associated with this fragmentation regardless of the deformation of the fragments.

All nuclear inputs to the SPY model, {\it i.e.} the individual binding energies, proton distributions and proton and neutron single-particle level schemes of each fission fragment as a function of its axial deformation are estimated for about 7000 nuclei from $Z = 20$ to $Z = 100$ within the framework of the constrained self-consistent Hartree-Fock-Bogoliubov (HFB) formalism using the Skyrme BSk27 effective nucleon-nucleon interaction \cite{goriely2013}. 
The nuclear state density is calculated in the framework of the statistical model of nuclear level densities \cite{lemaitre2019,DECOWSKI1968,MORETTO1972,MORETTO1972NPA,Demetriou01} on the basis of the discrete single-particle level scheme obtained self-consistently within the same HFB calculation.

\begin{figure}
\includegraphics[width=\columnwidth]{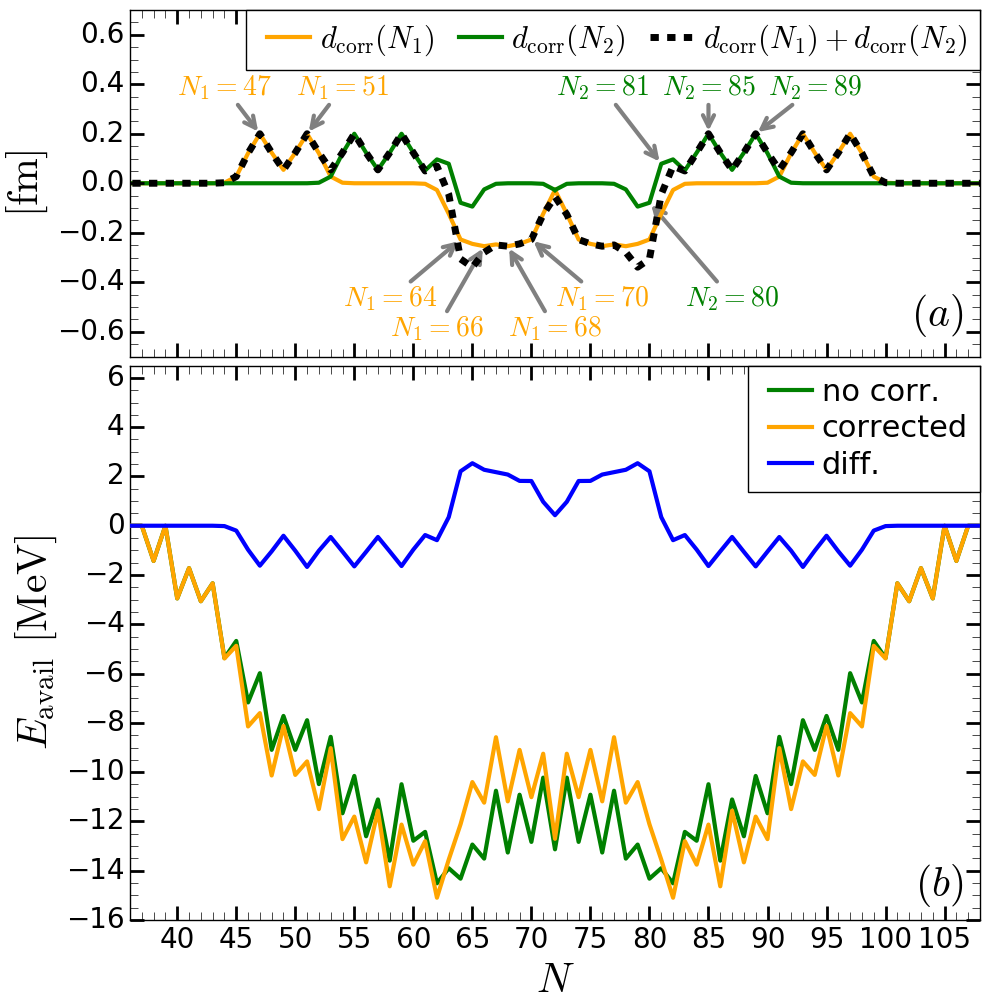}
\caption{(Color online) (a) Distance correction associated with the light fragment, $d(N_1)$ with $N_1\leqslant N_\mathrm{CN}/2$ (orange curve), with the heavy fragment, $d(N_2)$ with $N_2\geqslant N_\mathrm{CN}/2$ (green curve), and with the total one $d(N_1)+d(N_2)$ (black dashed curve) for $^{236}$U fission.
The specific neutron numbers $N_\mathrm{corr}$ (Eq.~\ref{eq:dcorr}) are indicated by gray arrows.
(b) Available energy of fragments from thermal neutron-induced fission of $^{235}$U calculated with (red curve) or without  (green curve) the phenomenological distance correction (Eq.~\ref{eq:dcorr}). The blue line corresponds to the the difference between corrected and non-corrected energies.}
\label{fig:AEU6}
\end{figure}

A scission criteria is needed to characterize the scission configurations, more precisely, to define the relative position of the fragments, hence to estimate the Coulomb and nuclear energies.
This criteria can be based on an inter-surface distance, as in the initial version of the SPY model \cite{lemaitre2015} or through the proton density at the scission neck which defines the scission distance as in our new version of the SPY model \cite{lemaitre2019}.
Presently, with respect to our last study \cite{lemaitre2019}, a  phenomenological modification of the scission distance is introduced in order to better reproduce experimental fission yields.
The corrected scission distance ($d_\mathrm{sc,corr}$) for a fragmentation $(Z_1,N_1)+(Z_2,N_2)$ is  defined through the proton density at the scission neck ($d_\mathrm{sc}$), as in Ref.~\cite{lemaitre2019}, though with two additional corrections, one per fragment
\begin{equation}
d_\mathrm{sc,corr}=d_\mathrm{sc}+d_\mathrm{corr}(N_1)+d_\mathrm{corr}(N_2)~.
\label{eq:dtot}
\end{equation}
Each correction term is composed by a positive term and a negative one which are chosen to be dependent on the difference between the fragment neutron number ($N$) and some specific neutron numbers ($N_\mathrm{corr}$), {\it i.e.} 
\begin{equation}
\label{eq:dcorr}
\begin{aligned}
d_\mathrm{corr}(N)=&\sum_{N_\mathrm{corr}\in\{47, 51, 81, 85, 89\}} 0.2~e^{-\frac{(N-N_\mathrm{corr})^2}{2}}\\
-&\sum_{N_\mathrm{corr}\in\{64, 66, 68, 70, 80\}} 0.2~e^{-\frac{(N-N_\mathrm{corr})^2}{2}} \ [\mathrm{fm}]~.
\end{aligned}
\end{equation}
Both light and heavy fragments contribute to the distance correction (Fig.~\ref{fig:AEU6}a) with a maximal correction of $\pm 0.4$~fm.
The available energy is impacted by a few MeV, for example by around 3~MeV for the neutron-induced fission of $^{235}$U (Fig.~\ref{fig:AEU6}b). 
Such an effect is of the same order of magnitude as pairing effects and can be regarded as a way to correct the approximate pairing effects which assumes the pairing gap at zero temperature can be described by a five-points mass difference \cite{Madland1988} and remains constant with the nucleus deformation.
For nucleosynthesis applications, the distance correction is kept constant irrespective of the fissioning system, even if a specific adjustment could be performed.

\begin{figure*}
\includegraphics[width=\textwidth]{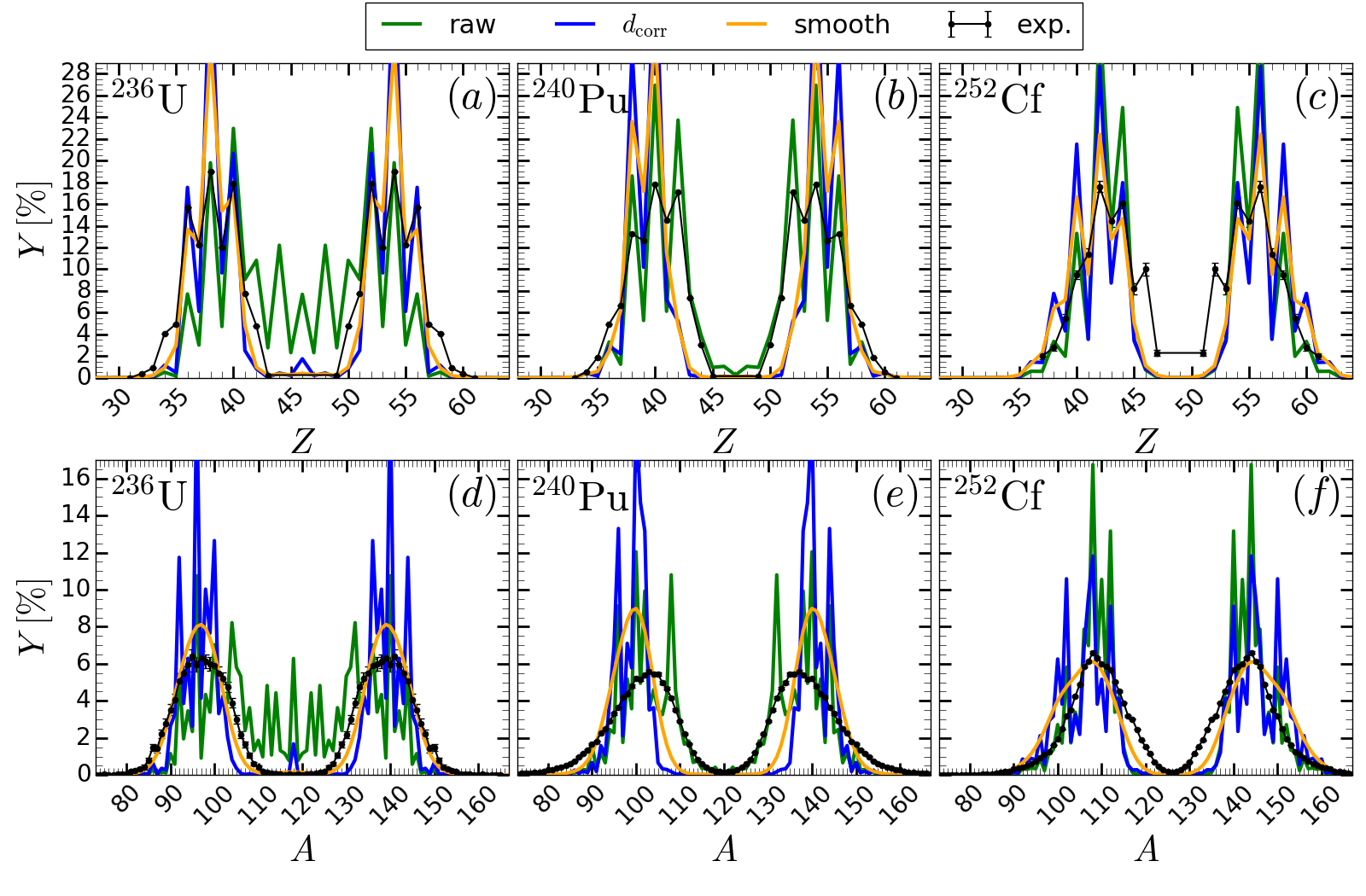}
\caption{
(Color online) Isotopic (top panels) and isobaric (bottom panels) fission yields of $^{236}$U, $^{240}$Pu and $^{252}$Cf.
The black lines with dots represent experimental (pre-neutron-emission) fission yields for fission of $^{236}\mathrm{U}$ $(Q=6.5~\mathrm{MeV})$ (a) \cite{SCHMITT1984} and (d) \cite{Romano2010}, $^{240}\mathrm{Pu}$ $(Q=6.5~\mathrm{MeV})$ (b) \cite{MARIOLOPOULOS1981} and (e) \cite{TSUCHIYA2000} and $^{252}\mathrm{Cf}$ $(Q=0~\mathrm{MeV})$ (c) \cite{BOCQUET1989} and (f) \cite{Zeynalov2011}. 
The green lines (labeled by ``raw") correspond to raw fission yields, the blue lines (labeled by ``$d_\mathrm{corr}$") to raw yields including distance corrections and the orange lines (labeled by ``smooth") to the final smooth yields after distance corrections.}
\label{fig:YaUPuCf}

\includegraphics[width=\textwidth]{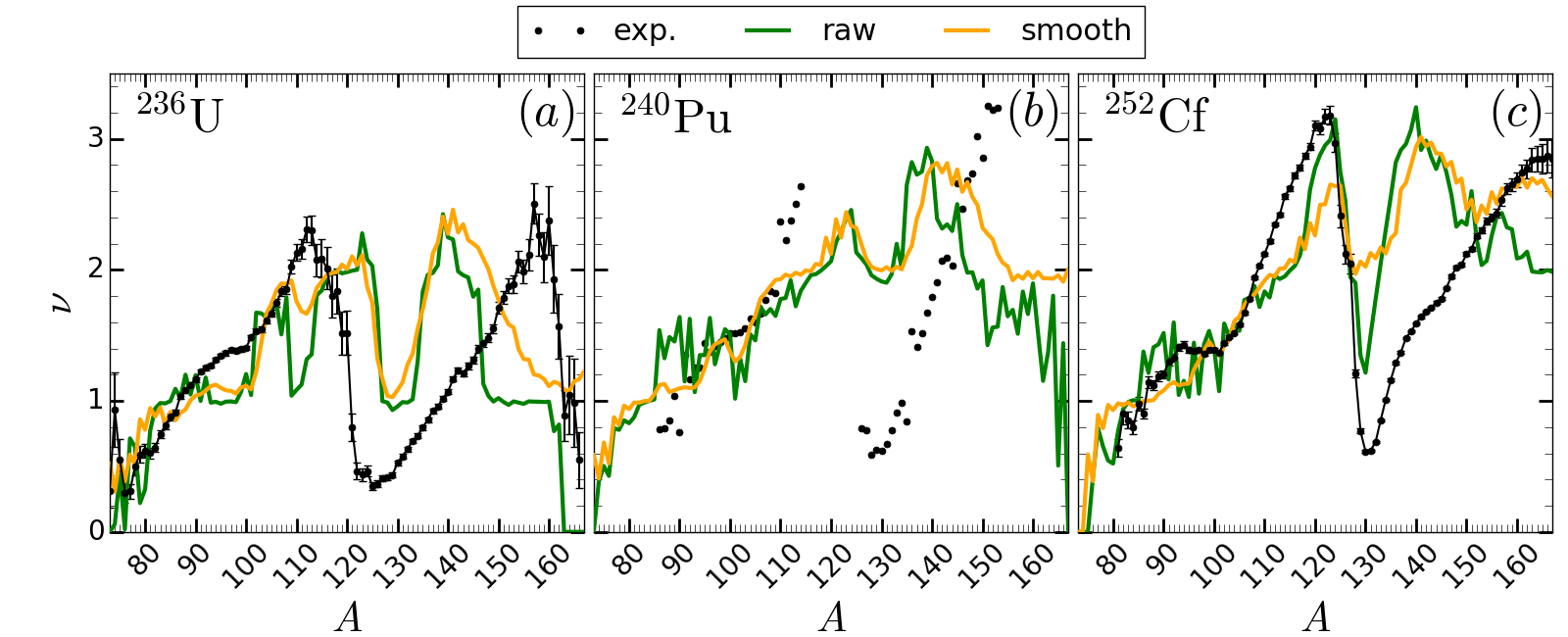}
\caption{
(Color online) Evaporated neutron distributions as a function of the fragment mass number. The black dotted lines represent experimental evaporated neutron distribution of (a) $^{236}\mathrm{U}$ $(Q=6.5~\mathrm{MeV})$ \cite{Vorobyev2010}, (b) $^{240}\mathrm{Pu}$ $(Q=6.5~\mathrm{MeV})$  \cite{NISHIO1995} and (c) $^{252}\mathrm{Cf}$  $(Q=0~\mathrm{MeV})$ \cite{DUSHIN2004}. 
The orange/green lines (labeled by ``smooth"/``raw") are SPY evaporated neutron distribution using smooth/raw pre-neutron yields.
}
\label{fig:NuaUPuCf}
		
\end{figure*}

This distance correction help to better reproduce the yields of the thermal neutron induced fission of $^{235}$U (Fig.~\ref{fig:YaUPuCf}a and d, green and blue curves), more particularly its symmetric component is reduced due to the distance correction around $N=64, 66, 68, 70$. 
As seen in Fig.~\ref{fig:YaUPuCf}c and f, this correction has a negligible impact on the FFD for the spontaneous fission of  $^{252}$Cf.
However, yields of the thermal neutron-induced fission of  $^{239}$Pu around $A=130$ becomes underestimated (Fig.~\ref{fig:YaUPuCf}b and e) due to the same correction as in the U case.

As described in Ref.~\cite{lemaitre2019}, the SPY FFD or yields distribution present strong staggering patterns that are not found experimentally. Those can be smoothed, as traditionally done, by a double Gaussian function
\begin{multline}
Y_\mathrm{smooth}(Z,N)=\\ \sum_{i=-4}^{4} \sum_{j=-15}^{15} Y_\mathrm{raw}(Z+i,N+j) C_z e^{-\frac{i^2}{2\sigma_z^2}} C_n e^{-\frac{j^2}{2\sigma_{n}^2}}
\label{eq:ysmoothed}
\end{multline}
where $C_z$ and $C_n$ are normalization factors and $\sigma_{z}=0.55$ and $\sigma_{n}=3$ are adopted.
In the present paper the width of the Gaussian function is larger in the neutron direction $\sigma_{n}$ than in the proton direction in order to be simultaneously consistent with experimental width of the isobaric and isotopic fission yields distributions (Fig.~\ref{fig:YaUPuCf}). 
As there is not coupling between $Z$ and $N$ in the double Gaussian function (Eq.~\ref{eq:ysmoothed}), the isotopic yields are not affected by the smoothing in the neutron direction.
Similarly to yields,  other fragments observables $\mathcal{O}$, such as the available energy or the kinetic energy, are also smoothed following the same procedure
\begin{multline}
\mathcal{O}_\mathrm{smooth}(Z,N)=\\ \sum_{i=-4}^{4} \sum_{j=-15}^{15} \mathcal{O}_\mathrm{raw}(Z+i,N+j)  C_z e^{-\frac{i^2}{2\sigma_z^2}} C_n e^{-\frac{j^2}{2\sigma_{n}^2}}.
\label{eq:OBSsmoothed}
\end{multline}

\subsection{Neutron evaporation model}
\label{sec:Nevap}

The SPY model allows us to investigate the primary fragments, \textit{i.e.} fragments formed at scission, hence their de-excitation by neutron evaporation.
The excitation energy of fragments has two components.
First, each fragment carries a fraction of the available energy at scission under an intrinsic excitation form.
Second, since each fragment can be deformed, a deformation energy may contribute to the excitation energy.
No assumption about available energy sorting at scission between primary fragments is needed within the SPY framework, contrary to other neutron evaporation models. Indeed, some models require to introduce an energy sorting parameter $R_T(A)$ \cite{Litaize2015fifrelin,Talou2011} or a thermal equilibrium to deduce the energy sorting on the basis of level density parameters with phenomenological corrections in order to increase the excitation energy of the light fragment \cite{VOGT2013}.

In the SPY model, the energy sorting, denoted by $x$ (Eq.~2 in Ref.~\cite{lemaitre2019}), results from the competition between the state density of the light fragment ($\rho_1$) and the heavy one ($\rho_2$).
The intrinsic excitation energy and deformation energy of each fragment are obtained by averaging over all possible energy sorting and deformations (Eq.~3 in Ref.~\cite{lemaitre2019}).
In the simplest model, the competition between neutron and gamma evaporation during the de-excitation cascade is not taken into account, so that the fission fragments are sequentially de-excited by neutron evaporation as long as it is energetically possible.
The kinetic energy of evaporated neutrons $E_\mathrm{n}$ can be assumed to follow a Maxwellian distribution
\begin{equation}
\phi_M(E_\mathrm{n})=\frac{2}{\sqrt{\pi T^3}}\sqrt{E_\mathrm{n}}e^{-E_\mathrm{n}/T},
\label{eq:maxwell}
\end{equation}
where $T$ is the nuclear temperature before neutron evaporation. $T$ can be extracted from the statistical model of nuclear level densities  including BCS pairing interaction, denoted as $T_\mathrm{BCS}$ (see Eqs.~16 and 18 in Ref.~\cite{lemaitre2019} and Refs.~\cite{DECOWSKI1968,MORETTO1972,MORETTO1972NPA}).

Following this approximation, the distribution of evaporated neutrons are calculated for the three well-known fissioning systems using both raw and smooth pre-neutron yields and shown in Fig.~\ref{fig:NuaUPuCf}.
In both cases, the number of emitted neutrons is underestimated for the heaviest fragments $A=150-160$ due to the low available energy, which, in turn, originates from an overestimate of the kinetic energy of these related fragmentations, as mentioned in Sec.~IV of Ref.~\cite{lemaitre2019}. 
However, for heavy fragments with $A=130-140$, the emitted neutron number is overestimated which, in this case, is due to an underestimate of the total kinetic energy of the fission fragments.
The distributions of evaporated neutron deduced from smooth yields (Fig.~\ref{fig:NuaUPuCf}, orange curves) follow roughly the same pattern as the raw ones except for the heaviest fragments region where the number of emitted neutron is larger.
This overestimation is due to the smoothing procedure (Eq.~\ref{eq:OBSsmoothed}) which leads to an increase of the total excitation energy by around 2~MeV, hence sometimes makes  the emission of an additional neutron possible depending on its kinetic energy.
More generally, the number of neutrons emitted  by the light fragment is in better agreement with data, though slightly underestimated, than for the heavy fragment where $\nu$ is systematically overestimated.
This is linked to the available energy sorting within a fragmentation which is too favorable for the heavy fragment. 

The overestimation the number of neutron emitted by the heavy fragments leads an overestimation of the total mean number of evaporated neutrons per fission.
As displayed in Table~\ref{tab:nubartot}, it is overestimated by 0.77-1.1 neutron when smooth yields are used and by 0.6-1 neutron for raw yields. 
The overestimation is higher with smooth yields due to a more important contribution of the heavy fragments.

\begin{table}[h]
	\caption{Total mean number of evaporated neutrons per fission, $\bar{\nu}_\mathrm{tot}$, from the evaluated data library ENDF/B-VII.1 \cite{ENDFB7} and  SPY using the raw or smooth yields (see Fig.~\ref{fig:NuaUPuCf}) for the 3 benchmarked nuclei shown in Fig.~\ref{fig:NuaUPuCf} and obtained either by neutron-induced fission (nif) or spontaneous fission (sf).}
    \begin{tabular}{ccccc}
  	\hline\hline
 	  Nucleus & reaction & ENDF & SPY(raw) & SPY(smooth) \\
 	 \hline  
 	 $^{235}\mathrm{U}$ & nif  & 2.44 & 3.08 & 3.20 \\
 	 $^{239}\mathrm{Pu}$ & nif & 2.88 &  3.97 & 3.96 \\
 	 $^{252}\mathrm{Cf}$ & sf & 3.77 & 4.37 & 4.39 \\
 	 \hline \hline
	\end{tabular}	
	\label{tab:nubartot}
\end{table}

To study the sensitivity of the predicted number of evaporated neutrons to the various model uncertainties, the impact of the neutron kinetic energy distribution $\phi(E_\mathrm{n})$, the nuclear temperature and the neutron-gamma competition are analyzed in the next sub-sections for the specific case of $^{252}$Cf spontaneous fission.

\subsection{Sensitivity studies of the evaporation model}
\label{sec:sensi_evapo}
\subsubsection{Impact of the kinetic energy distribution of the neutron}

\begin{figure}
\includegraphics[width=\columnwidth]{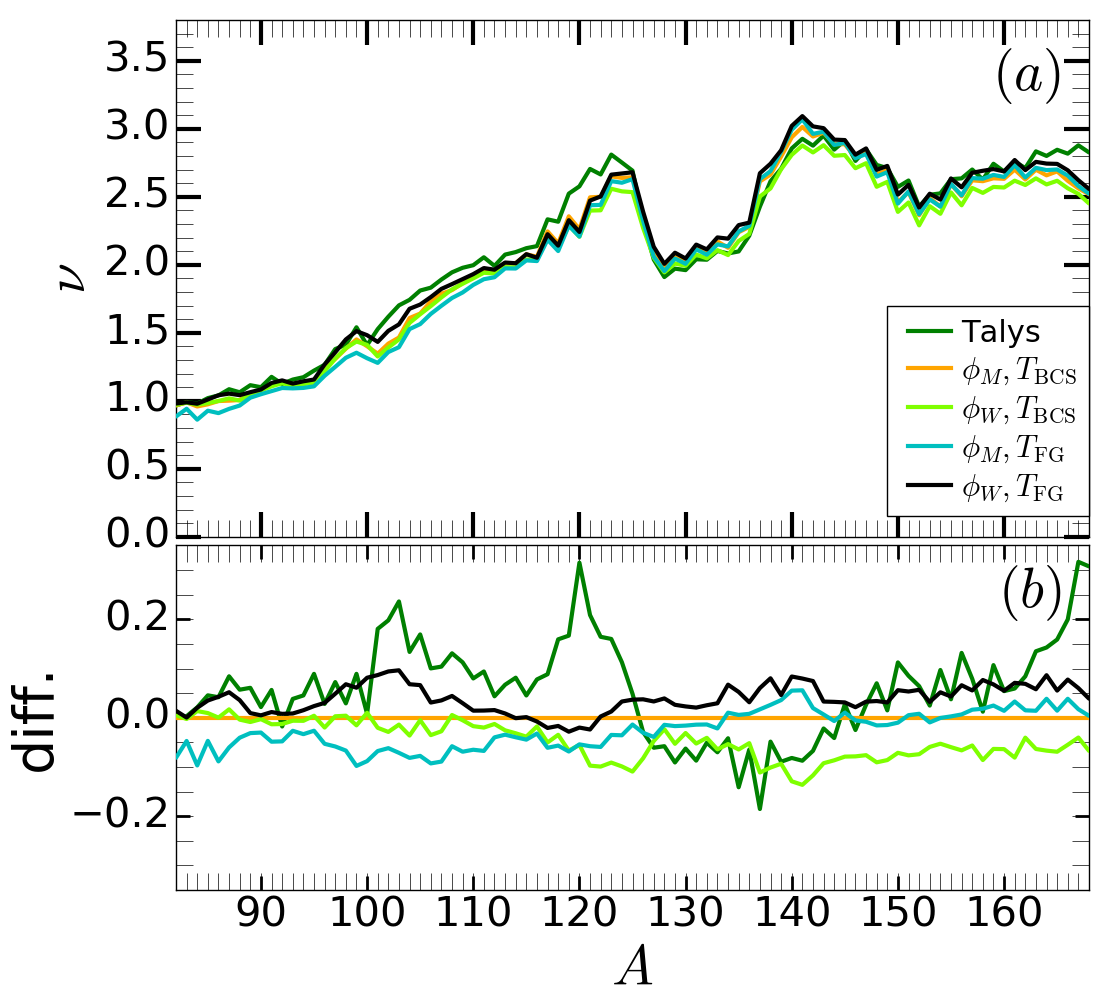}
\caption{(Color online) 
(a) Evaporated neutron distribution of spontaneous fission of $^{252}\mathrm{Cf}$ using the SPY raw yields. 
(b) Difference with respect to the evaporated neutron distribution computed with the Maxwell distribution (Eq.~\ref{eq:maxwell}) and temperature from the BCS state density.
}
\label{fig:NuaCfallopt}
\end{figure}

There are various models for the kinetic energy distribution of the neutron $\phi(E_\mathrm{n})$ \cite{KAWANO2013}.
The chosen one is related to the Maxwell-Boltzmann distribution describing the distribution of speeds of particles in an idealized gas which yields to the kinetic energy distribution of neutron $\phi_M$ (Eq.~\ref{eq:maxwell}).
Another possible description relies on the evaporation model of Weisskopf and Ewing \cite{Weisskopf1937,Weisskopf1940} yielding to
\begin{equation}
\phi_W(E_\mathrm{n})=\frac{E_\mathrm{n}}{T_\mathrm{n}^2}e^{-E_\mathrm{n}/T_\mathrm{n}}
\label{eq:weiss}
\end{equation}
for small $E_\mathrm{n}$ \cite{KAWANO2013} and where $T_\mathrm{n}$ is the nuclear temperature after the neutron emission.
The impact of the neutron kinetic energy distribution on the number of evaporated neutron of $^{252}$Cf spontaneous fission is negligible, as shown in Fig.~\ref{fig:NuaCfallopt}, (orange and light green curves, labeled respectively by ``$\phi_M,T_\mathrm{BCS}$" and  ``$\phi_W,T_\mathrm{BCS}$") where the difference is seen to rarely exceed 0.1 neutron (Fig.~\ref{fig:NuaCfallopt}b, light green curve, labeled by ``$\phi_W,T_\mathrm{BCS}$").

\subsubsection{Impact of the nucleus temperature}

As an alternative to the BCS theory of nuclear level densities, the phenomenological Fermi Gas approximation  \cite{Bethe1937},  excluding shell and pairing effects, can also be considered to estimate the nuclear temperature,
\begin{equation}
T_\mathrm{FG}=\sqrt{U/a}
\label{eq:Tfg}
\end{equation}
where $U$ is the excitation energy and $a\sim A/8$ is the empirical high-energy limit of the level density parameter \cite{TOKE1981}.
As seen in Fig.~\ref{fig:NuaCfallopt} (cyan and black curves, labeled respectively by ``$\phi_M,T_\mathrm{FG}$" and ``$\phi_W,T_\mathrm{FG}$"), both models for the nuclear temperature ($T_\mathrm{BCS}$ or $T_\mathrm{FG}$) give rather similar predictions of the evaporated neutron distribution. For the sake of coherence, $T_\mathrm{BCS}$ is adopted in the subsequent study.

\subsubsection{Impact of neutron-gamma competition}

In the previous calculation of the number of emitted number of neutrons, it was assumed that neutron emission dominates over the electromagnetic de-excitation. 
To test the impact of such an approximation, TALYS nuclear reaction code \cite{TALYS_KONING2012} is now used to describe the competition between the strong and electromagnetic channels in the de-excitation of the the primary fragments (Fig.~\ref{fig:NuaCfallopt}, green curve, labeled by ``TALYS"). 
Note that all de-excitation channels are taken into account in the Hauser-Feshbach and pre-equilibrium framework.
The difference between the evaporated neutron distribution from TALYS and the one obtained with the simplified Maxwellian approach described above is shown in Fig.~\ref{fig:NuaCfallopt} (green curve, labeled by ``TALYS") and does not exceed 0.3 neutron.

These results indicate that the primary fragments de-excitation can be essentially approximated by a sequence of neutron emissions, followed finally by gamma emission when no more neutron can be evaporated.
This simplified description based on energetic considerations seems to be sufficient within an accuracy of about 0.1--0.2 neutron.
This conclusion is only valid for low excited primary fragments up to 20--25~MeV.
The crucial point to improve the description of the emitted neutron distribution concerns the SPY model and particularly the kinetic energy of the fission fragments which directly affect the available energy and mean deformation of the fragments.

In the following, if not mentioned otherwise, the smooth yields (Eq.~\ref{eq:ysmoothed}) are adopted and neutron evaporation are deduced using the Maxwell distribution prescription (Eq.~\ref{eq:maxwell}) with the BCS temperature.

\begin{figure*}
  \subcaptionbox{Peak multiplicity for the raw pre-neutron isobaric yields with corrected distance (Eq.~\ref{eq:dtot}).\label{fig:syspic_rawd}}
  {  
  \includegraphics[width=\columnwidth]{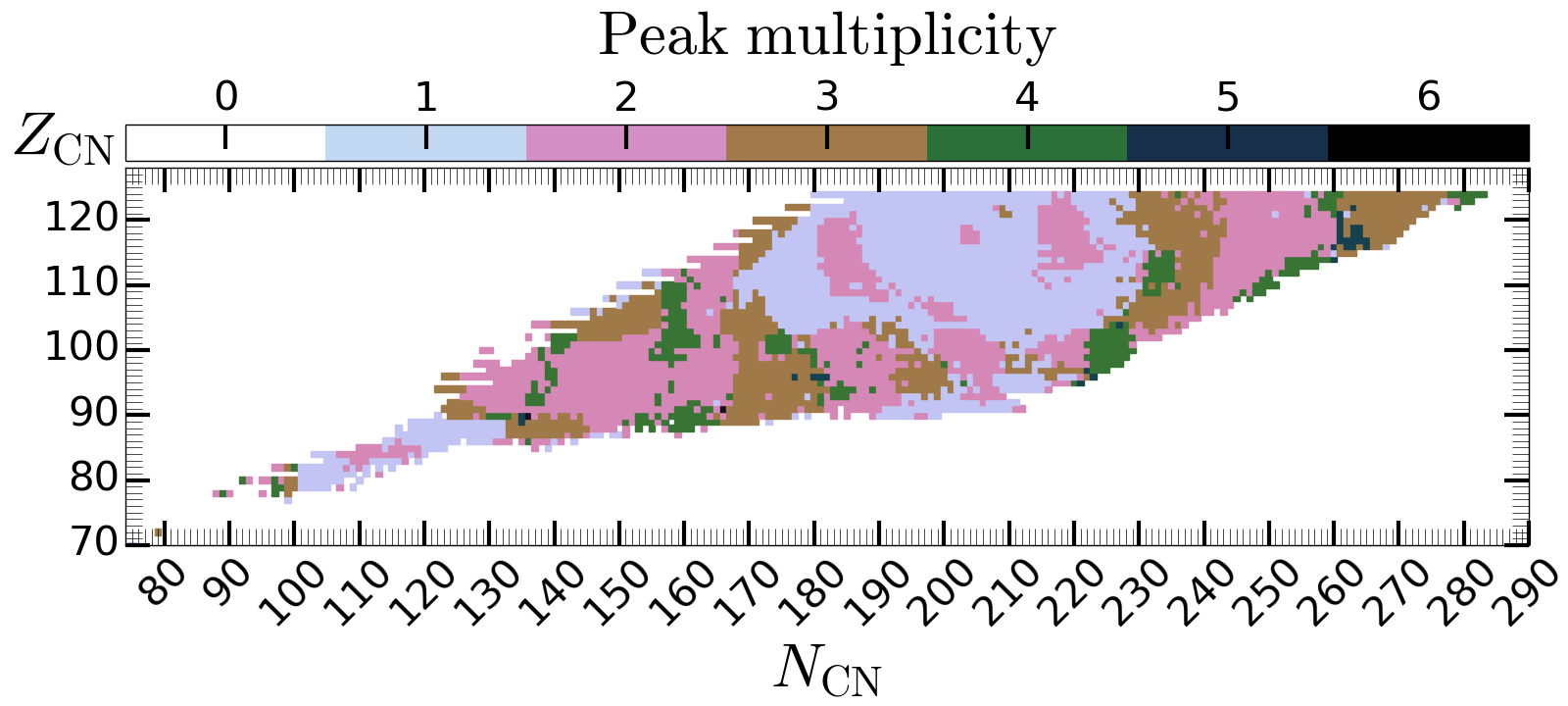}
  }
  \subcaptionbox{Mean prompt neutron multiplicity per fission according to the prescription given in Sec.~\ref{sec:Nevap}, deduced from the smooth pre-neutron isobaric yields with corrected distance (Eq.~\ref{eq:dtot}).\label{fig:sysnu}}
  {  
  \includegraphics[width=\columnwidth]{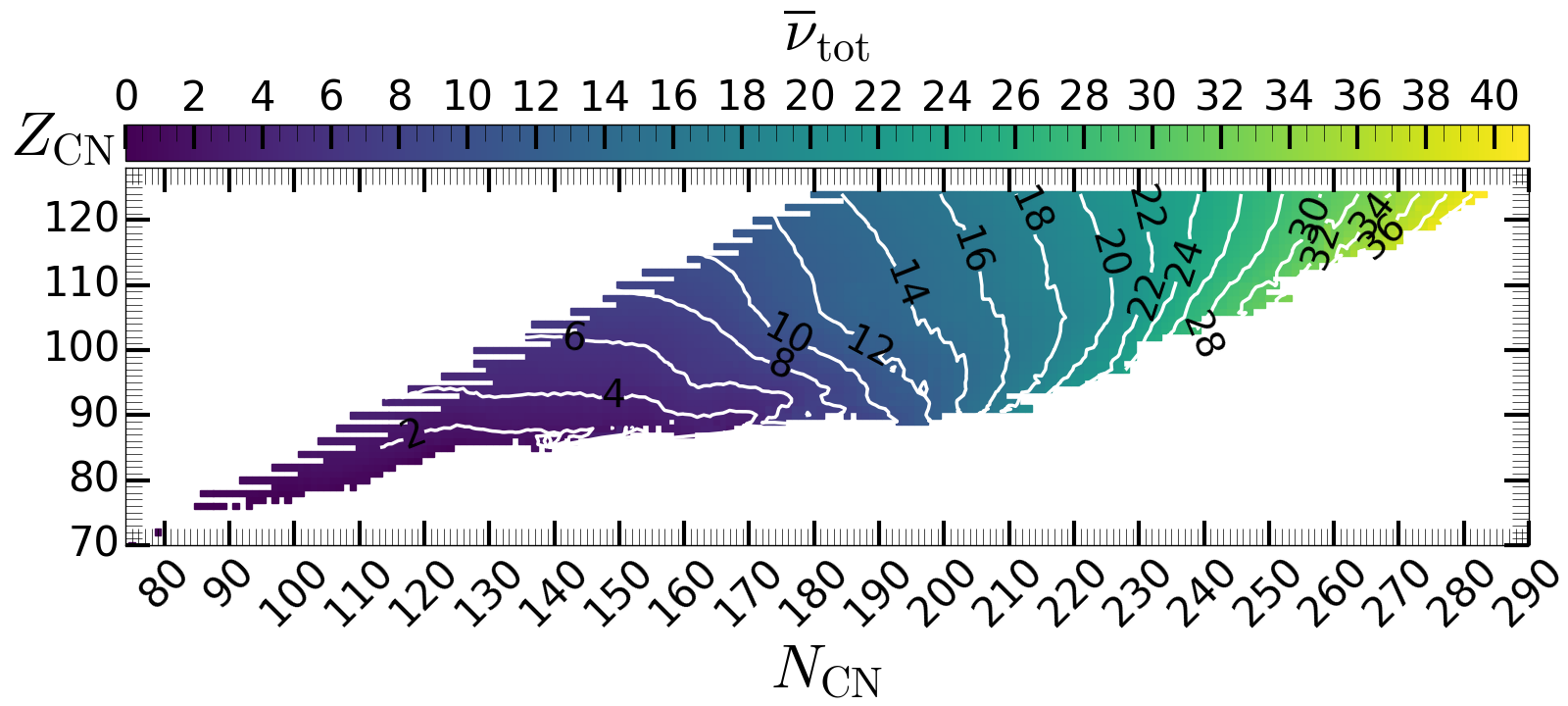}
  }   
  \hfill 
  \subcaptionbox{Peak multiplicity for the smooth pre-neutron isobaric yields with corrected distance (Eq.~\ref{eq:dtot}).\label{fig:syspic_smoothd}}
  {  
  \includegraphics[width=\columnwidth]{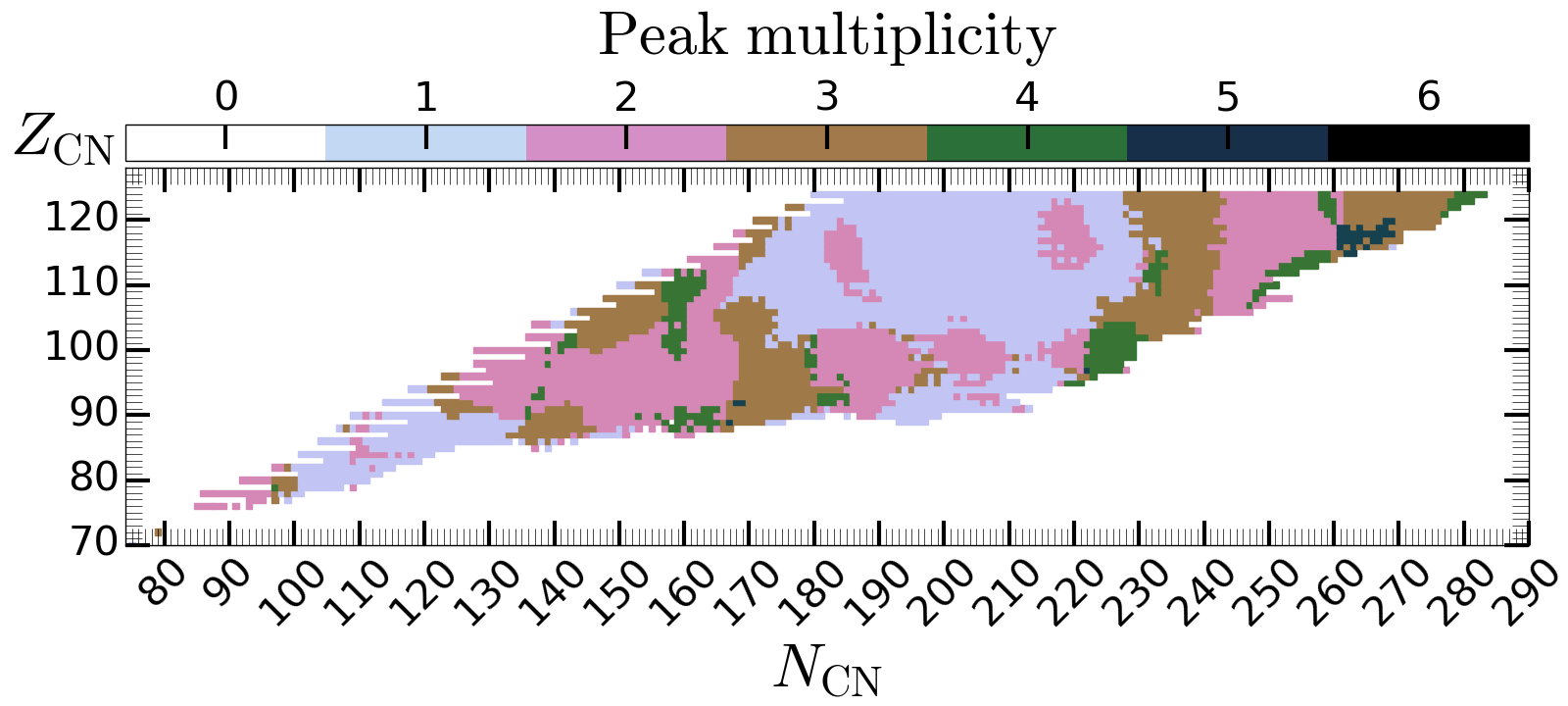}
  }
  \subcaptionbox{Mean available energy release per fission deduced from the smooth pre-neutron isobaric yields with corrected distance (Eq.~\ref{eq:dtot}).\label{fig:sysae}}
  {  
  \includegraphics[width=\columnwidth]{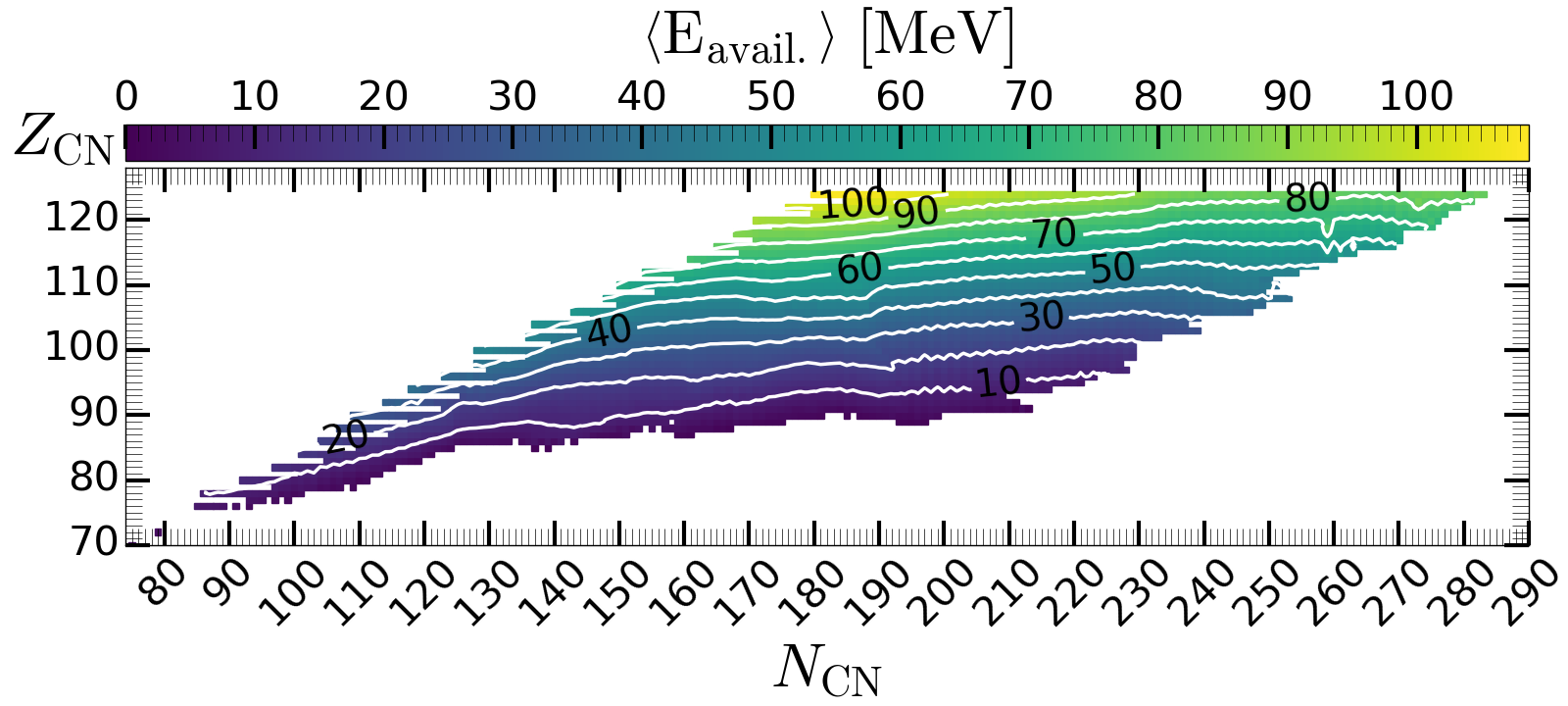}
  }
  \hfill 
  \subcaptionbox{Peak multiplicity for the smooth pre-neutron isobaric yields without corrected distance.\label{fig:syspic_smooth}}
  {  
  \includegraphics[width=\columnwidth]{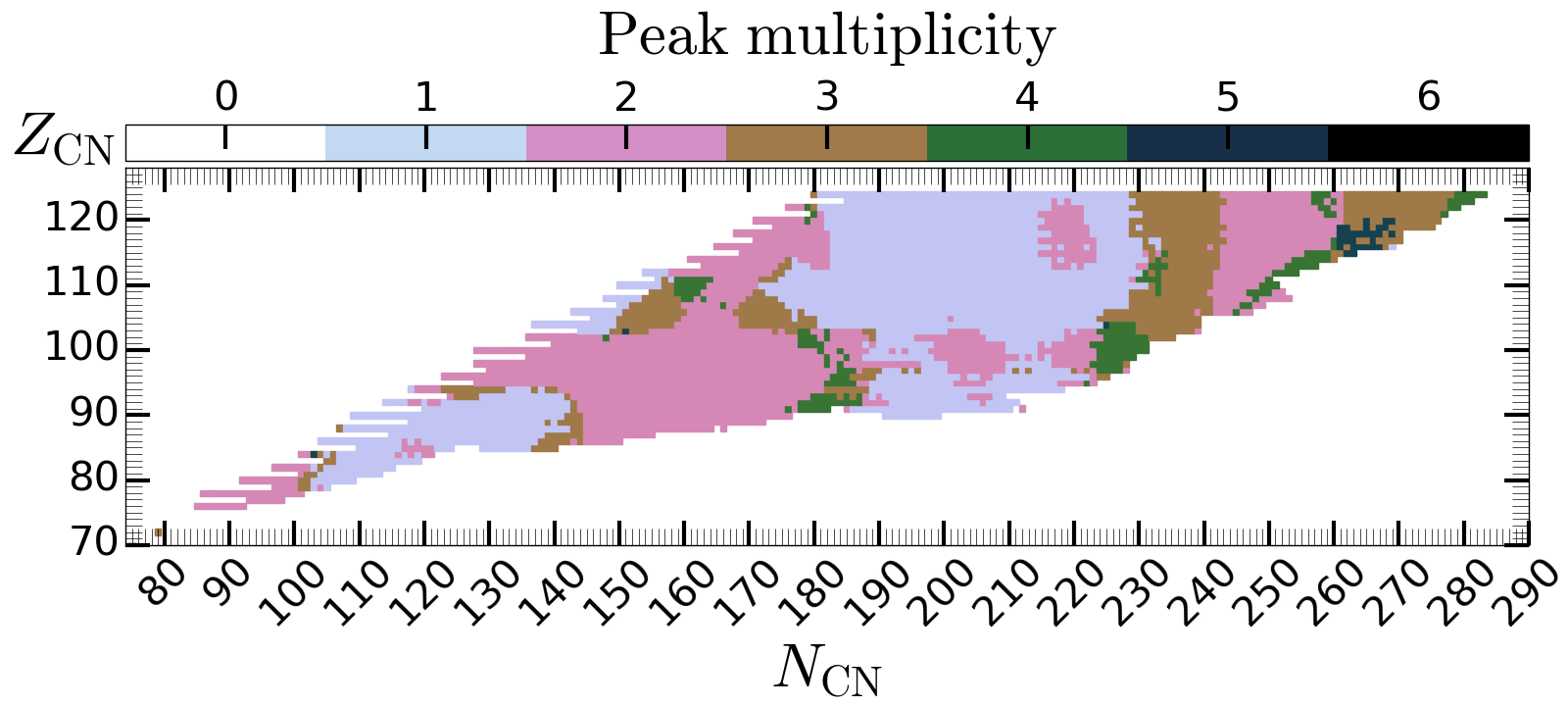}
  }
  \subcaptionbox{Isolines of mean available energy release per fission (orange lines) from Fig.~\ref{fig:sysae} and mean prompt neutron multiplicity per fission (green lines) from Fig.~\ref{fig:sysnu}.\label{fig:AEnu}}
  {  
  \includegraphics[width=\columnwidth]{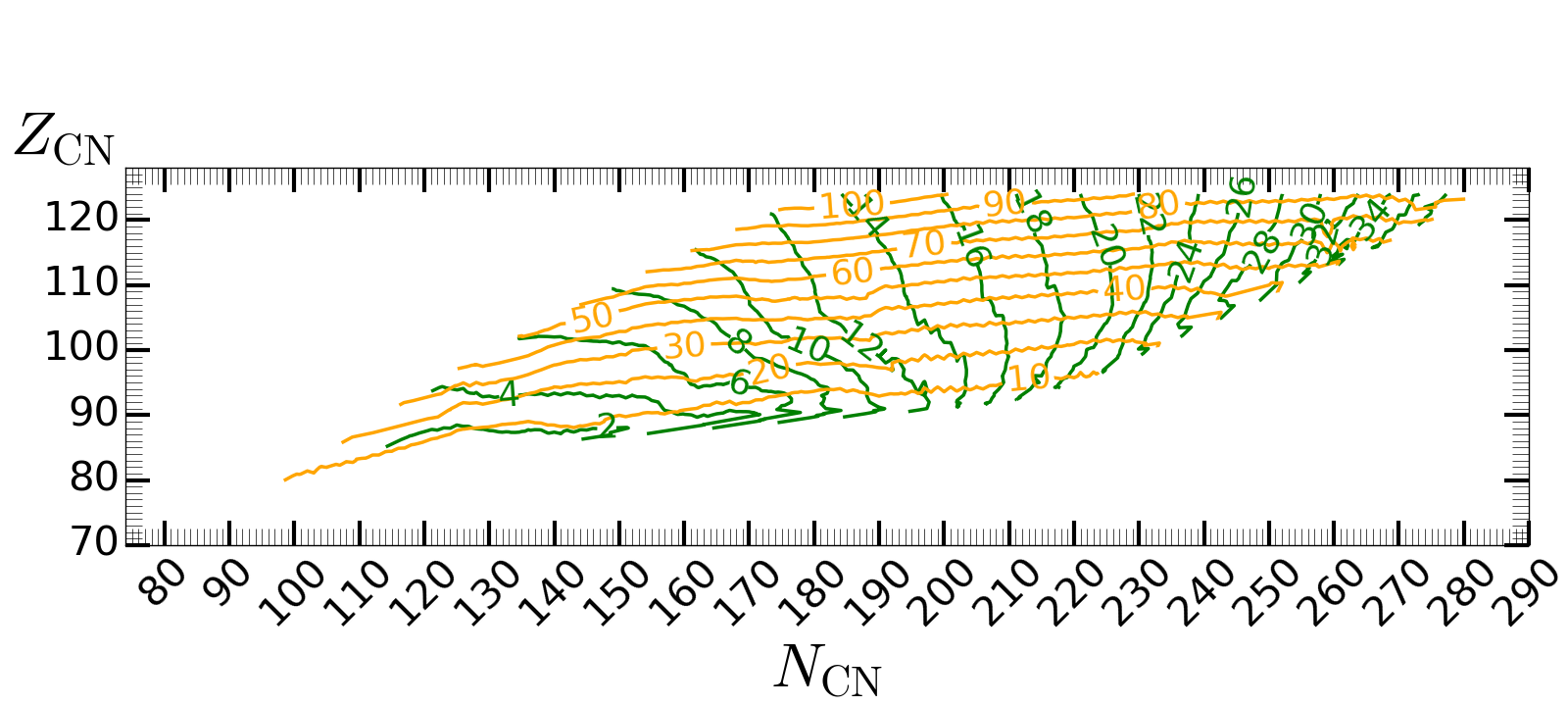}
  }
  
  \caption{(Color online) Systematics in the ($N,Z$) plane of major fission observables for some 3000 nuclei with $70 \leqslant Z \leqslant 124$ for an initial excitation energy of $Q=8\ \mathrm{MeV}$.}
  \label{fig:sys}
\end{figure*}

\subsection{Systematics}
\label{sec:sys}

\subsubsection{Over all fissionable nuclei}

Systematic SPY calculations were performed for isotopic chains between $Z=70$ and 124 from the proton to the neutron driplines with an initial excitation energy $Q=8\ \mathrm{MeV}$ (Fig.~\ref{fig:sys}). 
With $Q=8~\mathrm{MeV}$, only few nuclei with $Z<80$ can fission.
The smoothing procedure (Eq.~\ref{eq:ysmoothed}) may impact locally the peak multiplicity (Fig.~\ref{fig:syspic_smoothd}) when compared those obtained with raw yields (Fig.~\ref{fig:syspic_rawd}). 
In particular, peaks can be merged if there are close each other or if they are small enough.
However, the location of transition between the various fission modes are seen not to be globally affected by the smoothing procedure and does not distort significantly the peaks location of FFDs and their widths as shown in Fig.~\ref{fig:YaUPuCf}.

Four zones can be identified where the peak multiplicity is affected by the distance correction (Eq.~\ref{eq:dcorr}), as seen when comparing Figs.~\ref{fig:syspic_smoothd} and \ref{fig:syspic_smooth} obtained with or without this correction. In general terms, a negative/positive distance correction makes fragments closer/farther which induced a decrease/increase of the available energy, hence of the fission probability.
The first one is the shift of the symmetric/asymmetric fission transition toward neutron-deficient nuclei from $N_\mathrm{CN}\approx140$ to $N_\mathrm{CN}\approx133$.
The symmetric mode is disfavored by the negative distance correction for $N_\mathrm{frag}=64,66,68,70$ leading to the appearance of the asymmetric mode while for $N_\mathrm{frag}=47,51,81,85,89$ the positive distance correction favors the asymmetric mode.
In the second zone, located at $155 \lesssim N_\mathrm{CN}\lesssim 165$, a second asymmetric mode emerges, less asymmetric than the one without distance correction.
It is induced by a negative distance correction for the light fragment of the most asymmetric fission mode, {\it i.e.} $N_\mathrm{frag}=64,66,68,70$ and by a positive distance correction for the heavy fragment of the new slightly asymmetric fission mode ($N_\mathrm{frag}=85,89$).
The peak multiplicity increases from 2 to 4 when the distance corrections are important.
For more neutron-rich nuclei, in the third zone $165 \lesssim N_\mathrm{CN}\lesssim 180$, a symmetric mode emerges for high distance corrections.
It is induced by a negative distance correction due to the light fragment of the asymmetric fission mode ($N_\mathrm{frag}=64,66,68,70$) which disfavors the asymmetric mode.
The symmetric mode is further accentuated by the positive distance correction for $N_\mathrm{frag}=81,85,89$.
Finally, for nuclei in the fourth zone $180 \lesssim N_\mathrm{CN}\lesssim 200$, the positive distance correction for $N_\mathrm{frag}=85,89$ induces a slightly asymmetric fission mode on top of the symmetric one.
Although some fission yields distributions are impacted by the phenomenological correction to the scission distance, globally, SPY predictions for neutron-rich nuclei remains rather robust with respect to such a correction, as seen in Figs.~\ref{fig:syspic_smoothd} -- \ref{fig:syspic_smooth}.

The  mean number of neutrons  emitted per fission $\bar{\nu}_\mathrm{tot}$ ranges from 2 up to 37 neutrons depending on the fissioning system (Fig.~\ref{fig:sysnu}).
It depends mostly on the neutron richness of the fissioning nucleus. Its evolution with proton number is not trivial despite the strong variation of the mean available energy of fragmentations (Fig.~\ref{fig:sysae}).
For a given fissioning nucleus, an increase of the excitation energy ($Q$) enhances the intrinsic excitation energy of fission fragments, hence more neutrons are emitted. 
However, this relation between available energy and emitted neutrons is not sufficient to explain the relation between emitted neutrons and mean available energy of fragmentations.
Along a line of constant available energy (Fig.~\ref{fig:AEnu}) the number of evaporated neutron is multiplied by at least five between proton-rich and neutron-rich fissioning nuclei.
Similarly, along a constant line of emitted neutron number (Fig.~\ref{fig:AEnu}), the available energy increases by a factor up to hundred between light and heavy nuclei.
These features are due to the neutron binding energy of fission fragments, \textit{i.e.} the more neutron-rich the fissioning nucleus, the more-neutron rich the fission fragments and the lower their neutron binding energy, hence the more neutrons can be evaporated.
Even if the fission of a heavy nucleus releases more available energy compared to a light one, the number of evaporated neutrons can be similar because the neutron binding energies of fragments from heavy-nucleus fission is higher (less neutron-rich) than the fragments from the lighter one.

\subsubsection{SPY vs GEF}

\begin{figure*}
\centering
	\subcaptionbox{Evaluated isotopic yields (formed by Coulomb excitation) from Ref.~\cite{SCHMIDT2000} converted into isobaric yields using the UCD (Unchanged Charge Density) hypothesis. For ${}^{230}\mathrm{Th}$ post-neutron yields from ENDF/B-VII database \cite{ENDFB7}. \label{fig:spygefexp_8492}}
	{	  
	\includegraphics[width=\textwidth]{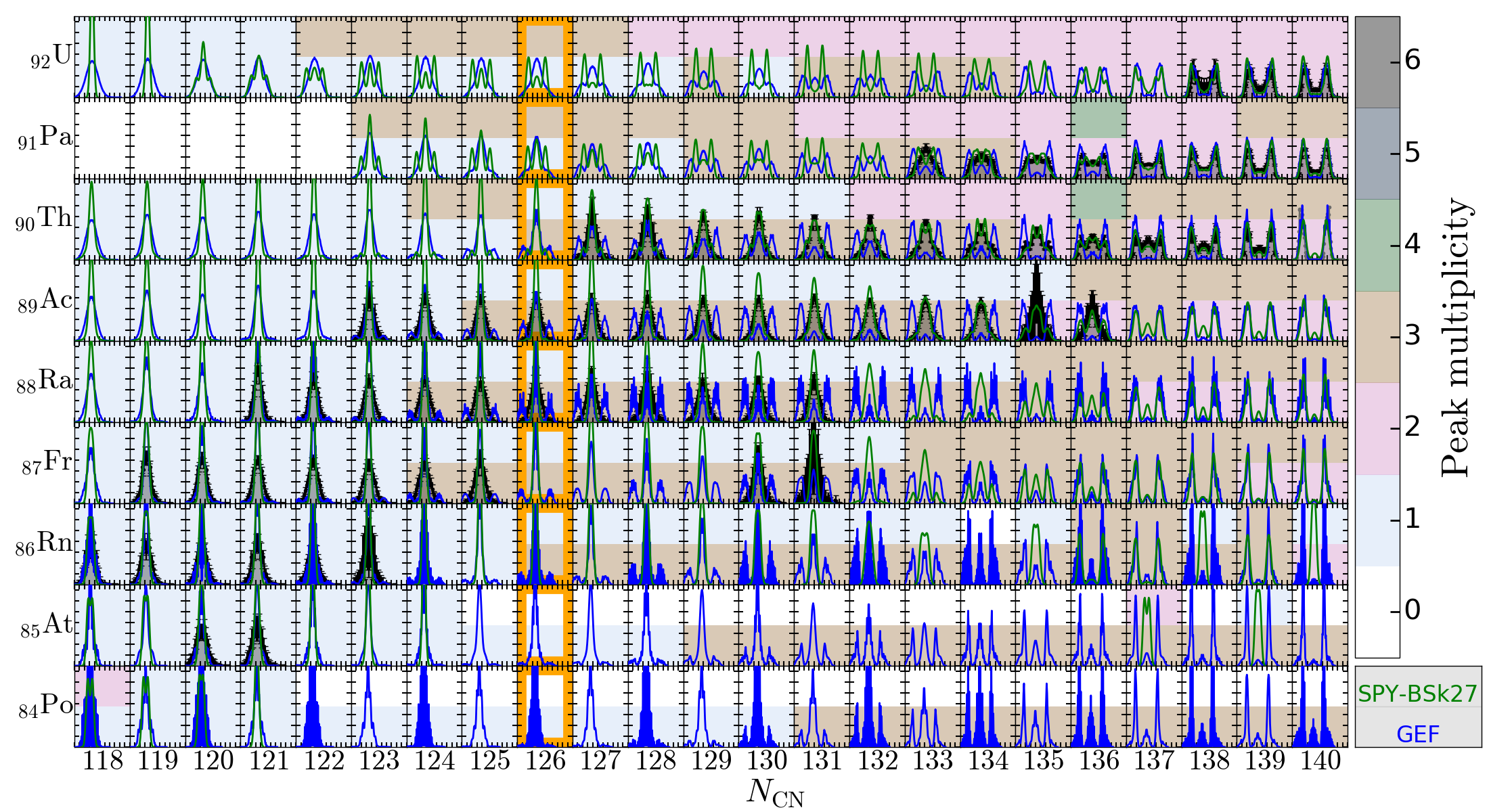}
	}
	\vfill
	\subcaptionbox{Yields of 
  ${}^{232-233}\mathrm{U}$ are from \cite{SCHMIDT2000} (converted into isobaric), 
  ${}^{234}\mathrm{U}$ from \cite{BABA1997}, 
  ${}^{235-237}\mathrm{U}$ (formed by 0.5~MeV neutron) from \cite{ENDFB7}, 
  ${}^{236}\mathrm{U}$ from \cite{Romano2010}, 
  ${}^{238}\mathrm{U}$ and ${}^{239}\mathrm{Np}$ and ${}^{244}\mathrm{Cm}$ from \cite{Ramos2018}, 
  ${}^{239}\mathrm{U}$ \cite{Zeynalov2011}, 
  ${}^{238}\mathrm{Np}$ from\cite{HAMBSCH2000}, 
  ${}^{240}\mathrm{Pu}$ from \cite{Thierens1981}, 
  ${}^{242-243}\mathrm{Pu}$ and ${}^{255}\mathrm{Es}$ (formed by thermal neutron) from \cite{ENDFB7}, 
  ${}^{243}\mathrm{Am}$ from \cite{RUBCHENYA2004}, 
  ${}^{244}\mathrm{Am}$ from \cite{Goverdovskij1995}, 
  ${}^{246}\mathrm{Cm}$ and ${}^{253}\mathrm{Es}$ and ${}^{254-256}\mathrm{Fm}$ (spontaneous fission) from \cite{ENDFB7}, 
 ${}^{248}\mathrm{Cm}$ from \cite{Zhao2003}, 
 ${}^{250}\mathrm{Cf}$ from \cite{unik1974}, 
 ${}^{252}\mathrm{Cf}$ from \cite{Zeynalov2011}, 
 ${}^{256}\mathrm{Cf}$ and ${}^{258}\mathrm{Fm}$ from \cite{Hoffman1980_21}, 
 ${}^{244-246-248}\mathrm{Fm}$ from \cite{Hoffman1980_22}, 
 ${}^{259}\mathrm{Lr}$ from \cite{Hamilton1992}, 
 ${}^{262}\mathrm{Rf}$ from \cite{Lane1996}.\label{fig:spygefexp_92104}}
	{	
	\includegraphics[width=\textwidth]{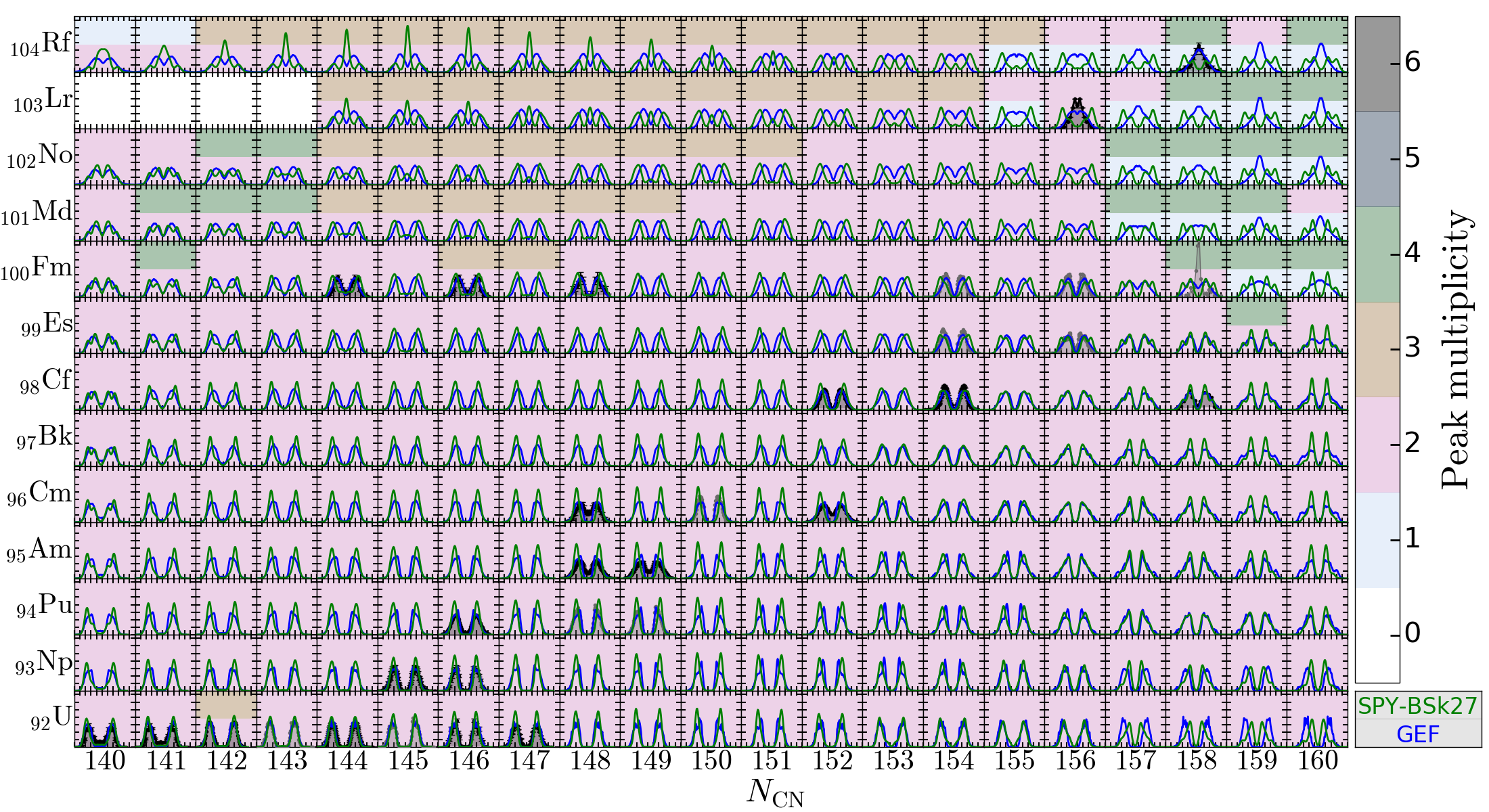}
	}

\caption{(Color online) Isobaric fission yields from SPY model (green curves) compared to those calculated by the GEF model (blue curves) \cite{SCHMIDT2016} and experimental pre-neutron (black curves) or post-neutron (gray curves) yields. For each FFD, the x axis (mass number) ranges from $A=70$ to $A=160$ by step of 10 and the y axis (FFD) from 0 to 15\% by step of 2\%. The background color refers to the peak multiplicity.}
\label{fig:spygefexp}
\end{figure*}

SPY FFDs with corrected distance are compared in Fig.~\ref{fig:spygefexp} to experimental data or with the evaluation from the ENDF/B-VII database \cite{ENDFB7}, if no experimental data is available.
Evaluated data (Fig.~\ref{fig:spygefexp}, gray curves) correspond to post-neutron FFDs while experimental data can be pre-neutron or post-neutron emission (Fig.~\ref{fig:spygefexp}, black curves).
SPY FFDs (Fig.~\ref{fig:spygefexp}, green curves) are also compared to GEF (version 2019/1.2) thermal neutron-induced fission yields \cite{SCHMIDT2016,gef2019}.
The excitation energies of experimental or evaluated data may not be exactly the same as those considered in SPY or GEF calculations but a variation of the excitation energy by a few~MeV only impacts moderately the shape of the FFD.

In the Po to U region (Fig.~\ref{fig:spygefexp_8492}), the transition between symmetric fission for neutron-deficient nuclei to asymmetric fission for neutron-rich nuclei is rather different between both SPY and GEF predictions.
SPY FFDs are in good agreement with experimental data \cite{SCHMIDT2000} for Th and Pa isotopes where the transition is well reproduced, in contrast to what is obtained by the GEF model.
The transition from symmetric to asymmetric fission predicted by GEF occurs for more neutron-deficient isotopes compared to experimental data.

For Th isotopes, GEF transition occurs between ${}^{216}\mathrm{Th}_{126}$ and ${}^{226}\mathrm{Th}_{136}$ for which no symmetric peak is found anymore.
From an experimental point of view, the transition seems to start at ${}^{222}\mathrm{Th}_{132}$ where the symmetric peak broadens while beyond ${}^{229}\mathrm{Th}_{139}$ the FFD is completely asymmetric.
The symmetric to asymmetric transition of Th predicted by SPY is slightly more complex (Fig.~\ref{fig:syspic_smooth} and Fig.~\ref{fig:spygefexp_8492}); there are two peaks in addition to the symmetric component for ${}^{222-226}\mathrm{Th}_{132-136}$ which gives a peak multiplicity of two and even four for ${}^{226}\mathrm{Th}_{136}$.
For more neutron-rich Th, this substructure disappears into one symmetric peak which become smaller for more neutron-rich isotopes and completely vanishes for ${}^{235}\mathrm{Th}_{145}$ (Fig.~\ref{fig:syspic_smooth}).
This kind of transition from symmetric to asymmetric through a substructure at the top of the vanishing symmetric peak is also visible on the SPY Pa FFDs.

The FFD broadening in the Rn to Ac isotopic chains is well reproduced by SPY even if the FFDs of neutron-deficient isotopes tend to be narrower than the experimental ones.
A strong odd-even staggering in GEF FFD is observed for even isotopes of Po and Rn which makes FFD plots completely blue filled (Fig.~\ref{fig:spygefexp_8492}).

For actinides from U to Rf (Fig.~\ref{fig:spygefexp_92104}), FFDs from SPY and GEF model are rather similar and in good agreement with experimental data, particularly for U and Es isotopes.
Compared to experimental data for Am and Cm isotopes, the symmetric part of the yields distributions is underestimated by both models.
The peaks are narrower with SPY compared to GEF model.
However, for neutron-rich nuclei (like ${}^{258}\mathrm{Fm}$), SPY predicts a purely asymmetric fission, contrary to GEF and evaluated data.
This asymmetric splitting predicted by SPY model is due to the doubly magic nuclei ${}^{132}\mathrm{Sn}$ which is disfavored.
Indeed, even if ${}^{132}\mathrm{Sn}$ has more available energy than other fragmentations, the available states for these fragmentations is significantly lower and consequently their yields is lower than other fragmentations leading to an asymmetric fission.
${}^{132}\mathrm{Sn}$ affects fissioning systems in the region around ${}^{258}\mathrm{Fm}$ (Fig.~\ref{fig:syspic_smooth}) where a doubly asymmetric fission is found by SPY model whereas GEF model predicts an symmetric mode.
For the same reasons, the symmetric fission of \textsuperscript{259}Lr and \textsuperscript{262}Rf are not reproduced with the SPY model contrary to the GEF model.

\begin{figure*}
\includegraphics[width=\textwidth]{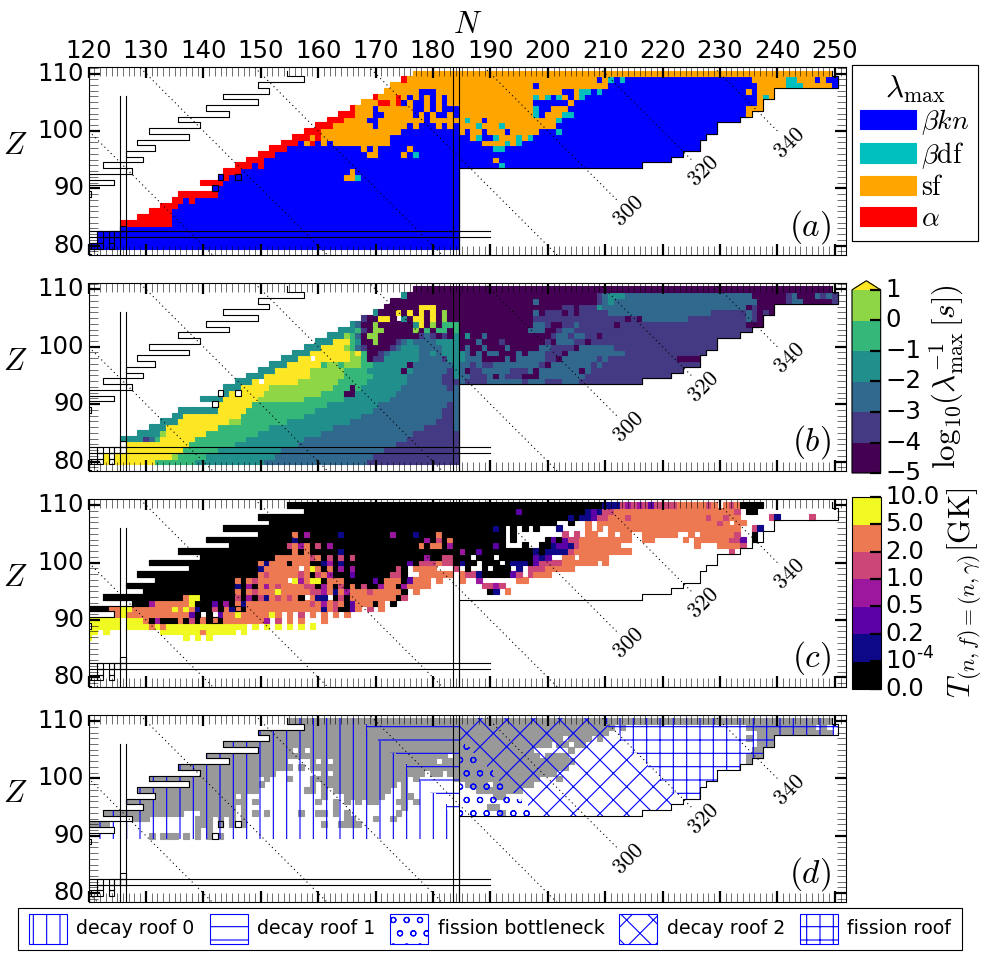}
\caption{
(Color online) Representation in the ($N$,$Z$) plan the decay mode with the highest decay rate (a) between $\beta$ decay followed by 0, 1, 2 or 3 neutron emission ($\beta kn$), $\beta$-delayed fission ($\beta$df), spontaneous fission (sf) or $\alpha$ decay; (b) the corresponding lowest lifetime $\tau_{\min}=1/\lambda_{\max}$ among the various decay mode given in (a). 
(c) Temperature at which fission and radiative neutron capture are equally favorable, above this temperature the neutron induced fission dominates the neutron capture. 
(d) Nuclei for which neutron-induced fission is dominant over the $(n,\gamma)$ channel for $T<0.2~\mathrm{GK}$ or for which the main decay mode is spontaneous fission or $\beta$-delayed fission are depicted by the gray area (region of efficient fission.
Blue hatched areas are the fission seeds areas~: the ``fission bottleneck" ($A<292$ and $N>184$) hatched by blue circles, the  ``fission roof" ($A\geqslant320$) hatched by blue vertical+horizontal lines, the ``decay roof 0" ($A<278$) hatched by blue vertical lines, the ``decay roof 1" ($A\geqslant 278$ and $N\leqslant 184$) hatched by blue horizontal lines and the ``decay roof 2" ($292\leqslant A<320$ and $N>184$) hatched by blue diagonal lines.
The double black lines show the $Z=82$, $N=126$ and $N=184$ shell closures. 
See text for more details. }
\label{fig:bknbdfsfa}
\end{figure*}

\section{Impact of fission on the r-process nucleosynthesis in NS mergers}
\label{sec:nucleo}

The NS-NS merger simulation was performed with a general relativistic Smoothed Particle Hydrodynamics scheme \cite{Oechslin07,Bauswein13,Bauswein10} representing the fluid by a set of particles with constant rest mass, the hydrodynamical properties of which were evolved according to Lagrangian hydrodynamics. 
We only consider dynamical (prompt) ejecta expelled during the binary NS-NS merger in the present paper since this ejecta component provides generally the most favorable conditions for fission recycling due to the low initial electron fraction.
A symmetric 1.365\Msun\ -- 1.365\Msun\ binary system compatible with the total mass detected in the GW170817 event \cite{Abbott17} has been modelled on the basis of the so-called SFHO equation of state \cite{Steiner2013}.
R-process calculations are performed on 500 ejected ``particles'' (\textit{i.e.} mass elements) representative of the dynamical ejecta.
The r-process nucleosynthesis is calculated with a reaction network including all 5000 species from protons up to $Z=110$ lying between the valley of $\beta$-stability and the neutron drip line. 
All charged-particle fusion reactions on light and medium-mass elements that play a role when the nuclear statistical equilibrium freezes out are included in addition to radiative neutron captures and photodisintegrations. 
The reaction rates on light species are taken from the NETGEN library, which includes all the latest compilations of experimentally determined reaction rates \cite{Xu13}. 
Experimentally unknown reactions are estimated with the TALYS code \cite{Goriely08a,TALYS_KONING2012} on the basis of the Skyrme HFB nuclear mass model, HFB-21 \cite{Goriely10a}. 
On top of these reactions, $\beta$-decays as well as $\beta$-delayed neutron emission probabilities are also included, the corresponding rates being taken from the relativistic mean-field model of Ref.~\cite{Marketin16}. 
Fission processes, including neutron-induced, spontaneous and $\beta$-delayed fission, are carefully introduced in the network together with the corresponding FFD. 
Fission barriers obtained with the BSk14 Skyrme interaction are adopted here to estimate fission probabilities \cite{Goriely07a,Goriely09b}. 
The new smooth FFDs from SPY model, as described in the previous section, as well as GEF predictions for comparison, are used in the present study.
Note that all fission products with a yield larger than typically $10^{-5}\%$ are linked to the parent fissioning nucleus in the network calculation, {\it i.e.} about 500 fission fragments are included for each fissioning nucleus. 
The neutron-rich fission fragments located outside the network, {\it i.e.} across the neutron drip line, are assumed to instantaneously emit neutrons down to the neutron drip line. The total mass fraction as well as number of nucleons is conserved at all time and the emitted neutrons are recaptured consistently by all the existing species. 
Finally, $\alpha$-decays are taken into account for all heavy species, the rates being extracted from Ref.~ \cite{Koura02}.

Despite the recent success of nucleosynthesis studies for NS mergers, the details of r-processing in these events is still affected by a variety of uncertainties.
In particular, the exact impact of neutrino is not yet understood in all details, the main reason of which is the not yet manageable computational complexity associated with full neutrino transport in a generically three-dimensional (3D), highly asymmetric environment with nearly relativistic fluid velocities and rapid changes in time. 
It was long assumed that neutrino interactions could not, at least not drastically, affect the initial neutron richness of the ejecta.
However, recent NS-NS merger simulations \cite{Wanajo14,Sekiguchi2015,Radice2016,Foucart2016,Lehner2016, Bovard2017,Martin2018,Ardevol19} including the effect of weak interactions on free nucleons demonstrate that neutrino reactions can significantly affect the neutron-to-proton ratio in the merger ejecta,
with direct consequences in particular on the amount of synthesized low-mass ($A<140$) r-nuclides produced in the dynamical ejecta along with the heavier species. 
To broadly study the impact of weak interactions of free nucleons on the r-process nucleosynthesis, a parametric approach in terms of prescribed neutrino luminosities and mean energies, guided by hydrodynamical simulations, was  followed in Ref.~\cite{Goriely15a}. 
A similar approach is also considered here to test the role of fission during the r-process nucleosynthesis in mainly two scenarios. 
The first one (hereafter scenario I) neglects all weak interactions on free nucleons and assumes the initial electron fraction of the ejected material is consequently not modified at the beginning of the expansion.
In this case, the mean initial electron fraction at the time of the network calculation ({\it i.e.} when the temperature has dropped below 10~GK) amounts to $\langle Y_e \rangle = 0.03$. This case is likely to be relevant for NS-BH mergers or highly asymmetric NS-NS mergers, where the lower mass component develops an extended tidal tail, from which considerable amounts of cold, unshocked matter can be centrifugally
ejected before neutrino exposure of these ejecta may play an important role \cite{Bauswein14,Just15,Foucart14}.
In the second case (scenario II), the equilibrium between electrons, positrons and neutrinos is assumed to hold down to the fiducial density of $\rho_\mathrm{eq}=10^{12}$\,g\,cm$^{-3}$, and from that density on, electron, positron and electron neutrino and antineutrino captures on nucleons 
\begin{equation}
\begin{aligned}
\nu_e + n \rightleftharpoons p + e^- \\
 \bar{\nu}_e + p \rightleftharpoons n + e^+ 
\end{aligned}
\label{eq:reac_neutri}
\end{equation}
are systematically included assuming the (anti)neutrino luminosities and mean energies remain constant in time, as detailed in Ref.~\cite{Goriely15a}, and compatible with the results obtained within the Improved Leakage-Equilibration-Absorption Scheme (ILEAS) method  \citep{Ardevol19} (see in particular their Fig. 14). 
To do so, we adopt the values of $L_{\nu_e} =0.3\times 10^{53}$~erg/s; $L_{{\bar\nu}_e}= 10^{53}$~erg/s; $\langle E_{\nu_e}\rangle=8$~MeV; $ \langle E_{{\bar\nu}_e}\rangle = 12$~MeV. The resulting mean electron fraction at the time of the network calculation reaches  $\langle Y_e \rangle = 0.23$ in relatively good agreement with the 0.26 value obtained in Ref.~\cite{Ardevol19}.

Without weak interaction,  the initial electron fraction, $Y_{e}^0$, of the various mass elements is low and characteristic of the NS inner crust, {\it i.e.} less than 0.06; consequently the number of free neutrons per seed nuclei can reach a few hundreds. 
With such a neutron richness, heavy fissioning nuclei can be efficiently produced during the r-process.
Weak interactions on nucleons (Eq.~\ref{eq:reac_neutri}) increase the initial electron fraction in the ejecta to values ranging in this scenario II from 0.06 to 0.38 and make the production of heavy fissioning nuclei and fission recycling during the r-process less favorable, though not impossible.

On the basis of the BSk14 fission barriers adopted here, the production of super-heavy neutron-rich nuclei is possible up to $Z=110$ if the neutron irradiation is large enough to overcome the $N=184$ shell closure.
Beyond $Z=110$, the production of elements is unlikely because these elements fission spontaneously (Fig.~\ref{fig:bknbdfsfa}a) with a very short lifetime (Fig.~\ref{fig:bknbdfsfa}b).
This is the main reason why the reaction network has been limited to elements $Z\leqslant 110$.
Nuclei reaching this limit ($Z=110$ and $A\geqslant320$) define a ``fission roof" area ($A\geqslant 320$) (Fig.~\ref{fig:bknbdfsfa}d).

To produce such super-heavy neutron-rich nuclei, the matter has to flow through the fissioning region formed by (\textit{i}) $\beta$-delayed fission of ${}^{286}_{94}\mathrm{Pu}_{192}$, (\textit{ii}) spontaneous fission of ${}^{286}_{95}\mathrm{Am}_{191}$, (\textit{iii}) $\beta$-delayed fission of ${}^{288-289}_{97}\mathrm{Bk}_{191-192}$, and (\textit{iv}) spontaneous fission of elements with $Z\geqslant 98$ (Fig.~\ref{fig:bknbdfsfa}a).
${}^{286}_{95}\mathrm{Am}_{191}$ and  $Z\geqslant 98$ elements have lifetime shorter than $10^{-4}$~s, which makes neutron capture by these nuclei unlikely.
The $\beta$-delayed fission lifetime of ${}^{286}_{94}\mathrm{Pu}_{192}$ and ${}^{288-289}_{97}\mathrm{Bk}_{191-192}$ ranges from $10^{-3}$ to $10^{-2}$~s.
The neutron-induced fission of ${}^{287-291}_{98}\mathrm{Cf}_{189-193}$ and ${}^{285}_{95}\mathrm{Am}_{190}$ isotopes also happens to be efficient and contributes to fission recycling (Fig.~\ref{fig:bknbdfsfa}c).
The flow through Pu isotopes is negligible due to the fast $\beta$-delayed neutron emission of ${}^{278}_{94}\mathrm{Pu}_{184}$ with a lifetime $\tau_{\beta kn}=6.6 \times 10^{-4}$~s and  ${}^{279}_{94}\mathrm{Pu}_{185}$ with $\tau_{\beta kn}=4.7 \times 10^{-3}$~s (Fig.~\ref{fig:bknbdfsfa}b).
In addition, the neutron separation energy of ${}^{279}\mathrm{Pu}$ equals to 0.28~MeV which makes the neutron capture by ${}^{278}\mathrm{Pu}$ unfavorable.
Consequently, matter mainly flows through the Cm isotopic chain to produce super-heavy neutron-rich elements.
This region forms a ``fission bottleneck" ($A<292$ and $N>184$) where ${}^{287}_{96}\mathrm{Cm}_{191}$ represents the ``crossing point'' (Fig.~\ref{fig:bknbdfsfa}d).
However, the production of super-heavy neutron-rich nuclei is no more possible for temperatures higher than 2~GK ({\it i.e.} $k_BT=0.17$~MeV) because the neutron-induced fissions become more favorable than the radiative neutron captures of ${}^{285-288}_{96}\mathrm{Cm}_{189-192}$ and ${}^{285}_{97}\mathrm{Bk}_{188}$ (Fig.~\ref{fig:bknbdfsfa}c) which close the ``fission bottleneck".

At the end of the neutron irradiation, heavy neutron-rich elements $\beta$-decay up to the region of efficient fission.
Elements accumulated during the neutron irradiation along the $N=184$ neutron shell closure $\beta$-decay up to reach a region of efficient fission.
This can be split in two zones~: the ``decay roof 0" ($A<278$) and ``decay roof 1" ($A \geqslant 278$ and $N \leqslant 184$) which correspond to the region reached by nuclei $Z<94$ and $Z\geqslant 94$, respectively.
Elements with $292\lesssim A \lesssim 320$ accumulated after the ``fission bottleneck" $\beta$-decay up to reach another region of efficient fission named ``decay roof 2" ($292\leqslant A<320$).

In the following subsection (Sec.~\ref{sec:finalabund}), we present the final abundances of the ejected material obtained in both scenarios as well as their sensitivity to the initial conditions of the trajectories.
The fission contribution to the r-process and its impact on the final abundances will be studied in both scenarios in Sec.~\ref{sec:FR}.
The contributions of the various fissioning regions to the fission recycling is studied in Sec.~\ref{sec:FS}, while, in Sec.~\ref{sec:FP}, the impact of the FFDs on the final abundances are studied with a specific comparison made when adopting SPY or GEF FFDs.

\subsection{Final abundances}
\label{sec:finalabund}

\begin{figure}
\includegraphics[width=\columnwidth]{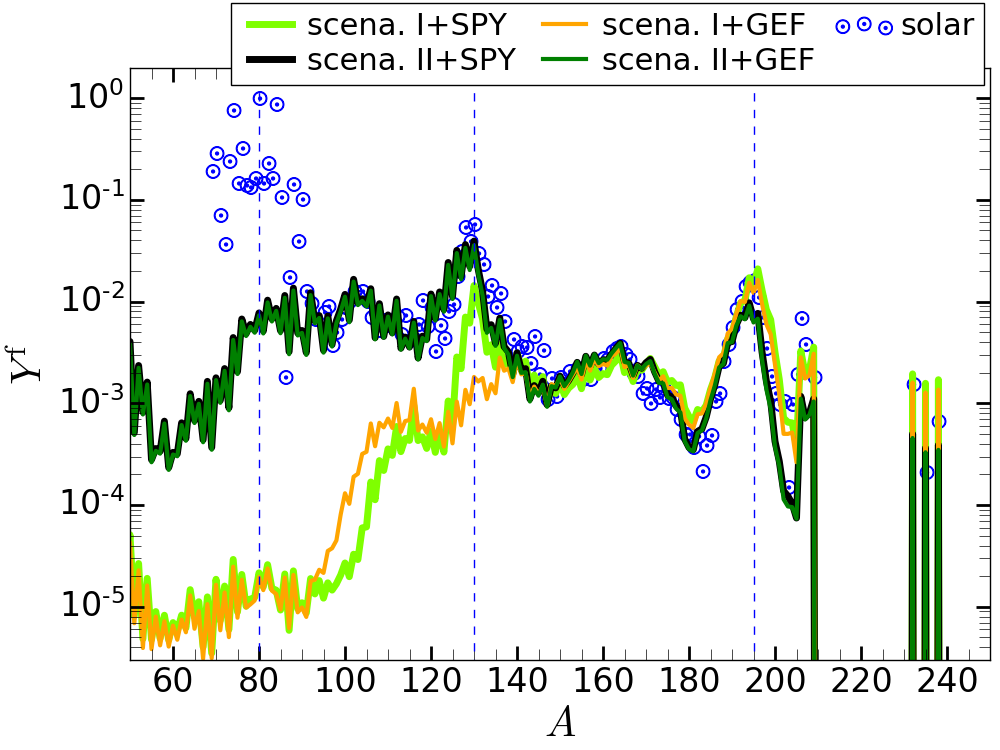}
\caption{
(Color online) Final composition (in terms of molar fraction) of matter ejected in scenarios~I and II and obtained either with SPY or GEF FFDs. For comparison, the solar r-abundance distribution \cite{Goriely1999} (blue dotted circles) is shown and arbitrarily normalized.
}
\label{fig:allmYf}
\end{figure}

\begin{figure*}
\includegraphics[width=\textwidth]
{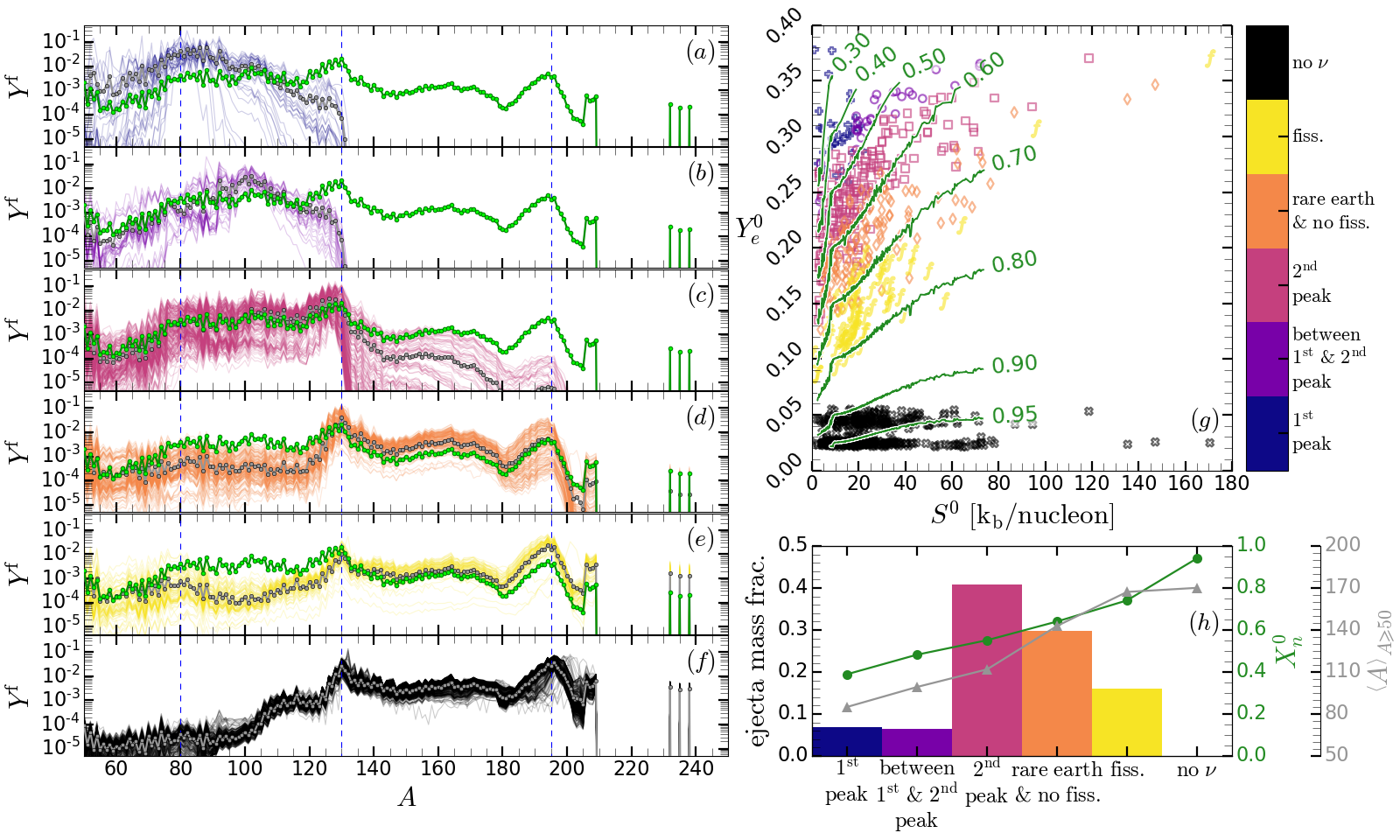}	
\caption{
(Color online) (a-e) Final abundances of the trajectories computed within scenario~II classified into 5 groups~: ``1\textsuperscript{st} peak", ``between 1\textsuperscript{st} and 2\textsuperscript{nd}peak", ``2\textsuperscript{nd} peak", ``rare earth \& no fiss." and ``fiss.".
The green curves with dots are the final abundances averaged over all trajectories computed within scenario~II (a-e).
The gray curves with dots correspond to the mean final abundances within each group.
(f) Final abundances of the trajectories computed within scenario~I (black curves) and the final abundances averaged over all trajectories (gray curve with dots).
(g) Distribution of the trajectories in the ($S^0$,$Y_e^0$) plane and iso-$X_n^0$ curves (green curves).
(h) Mean value of the initial neutron mass fraction $X_n^0$ (green curve) and mean mass $\left<A\right>_{A\geqslant 50}$ (gray curve) of each group and the associated ejecta mass fraction of groups for the scenario~II (color bars).
}
\label{fig:YAfSYe}
\end{figure*}

The final abundance distributions of the ejected material are shown in Fig.~\ref{fig:allmYf} for both scenarios I and II. 
Calculations are performed adopting either the SPY or GEF predictions of the FFDs. 
While in scenario I, only nuclei with $A\gsimeq 120$ are essentially produced,  weak interactions on nucleons also allow the production of lighter species with $80 \lsimeq A \lsimeq 120$ in scenario II, but also decrease the production of the Pb peak nuclei as well as the long-lived actinides.
In this case, the choice of the FFD model (SPY or GEF) is seen to have a negligible impact on the final abundance distribution. 
The same conclusions hold if we consider our original version of the SPY FFD \cite{lemaitre2019}.
Without weak interactions, with the low initial electron fractions, the number of neutron per seed is rather high (a few hundreds), so that the production of the heaviest nuclei is efficient.  
In this case, nuclear fission plays a fundamental role during both the neutron irradiation and at  freeze-out \cite{Goriely2015} and the FFD model directly affects the abundance distribution, particularly around the second peak $A=130$ where SPY FFDs give higher abundances than GEF (see Sec.~\ref{sec:FP}). 
This nucleosynthesis has been described in detail in Refs.~\cite{Bauswein13,Goriely11b,Goriely13c,Just15,Goriely15a,Goriely19b}. 
Note that the corrections introduced in the present study (Sec.~\ref{sec:spypres}) in comparison with our 2019 version \cite{lemaitre2019} have a small impact on the final abundance distribution. They tend to slightly increase the second r-abundance peak at $A\simeq 130$ and decrease its low-$A$ tail. Such differences are significantly smaller than the one observed between the SPY and GEF predictions.

It is well accepted \cite{Hoffman97} that the efficiency of the r-process nucleosynthesis can be related to three main physical properties of the ejected material, namely the electron fraction $Y_e$, the entropy per baryon $S$ and the expansion timescale $\tau$ \cite{Qian96}.
In scenario~I, trajectories are characterized by a low initial electron fraction $Y_e^0$ ranging between 0.02 and 0.06 at the beginning of the nucleosynthesis, \textit{i.e.} when the temperature has dropped below 10~GK and the density below the drip density ($\rho_\mathrm{drip}\simeq 4 \times 10^{11}$ g/cm\textsuperscript{3}) (Fig.~\ref{fig:YAfSYe}g). 
Initial entropies $S^0$ reach values up to $170~k_B/\mathrm{nucleon}$, though most of them remain below $50~k_B/\mathrm{nucleon}$. As already mentioned, in scenario~II, the initial electron fraction covers a much larger range between 0.06 and 0.38 due to the weak interactions (Eq.~\ref{eq:reac_neutri}).

Both the initial electron fraction and entropy affect the initial mass fraction of free neutrons $X_n^0$ (Fig.~\ref{fig:YAfSYe}g, green curves) which is consequently a parameter of first relevance in the analysis of the efficiency of the r-process nucleosynthesis. In scenario I, $X_n^0$ ranges between 0.85 and 0.98, as seen in Fig.~\ref{fig:YAfSYe}g, while in scenario II it varies between 0.02 and 0.82. The lower $Y_e^0$, the larger $X_n^0$ by definition, but also the higher the initial entropy, the larger $X_n^0$ due to the release of free neutrons through photodissociation. Therefore, $X_n^0$ engulfs information on both the initial electron fraction and entropy.
The sudden variation of $X_n^0$ observed at $S^0=8\mathrm{-}10~k_B/\mathrm{nucleon}$, independently of $Y^0_e$, is linked to the initial abundances of trans-alpha elements ($A>4$)~: $\sum_{A>4}Y^0(A)$.
The matter is not completely dissociated for trajectories with a low initial entropy $S^0<10~k_B/\mathrm{nucleon}$, which makes the initial abundance of trans-alpha elements (with a mean mass $\left<A\right>_{A>4}^0$ ranging from 50 to 85)  non-negligible and reduces the initial neutron mass fraction.
For $S^0>10~k_B/\mathrm{nucleon}$, the initial abundances of trans-alpha elements is negligible, {\it i.e.} the ejected nuclear matter is completely dissociated into protons, neutrons and alphas.
The low entropy effect is more important for trajectories in scenario~II where for $18\%$ of the ejected mass, $\sum_{A>4}Y^0(A)>0.001$, whereas in scenario~I, only $5\%$ of the ejecta has $\sum_{A>4}Y^0(A)>0.001$.
Trans-alpha elements can boost the r-process depending on the neutron irradiation.
Initial and final abundance distributions are more or less similar for trajectories with a low initial neutron mass fraction ($X_n^0<0.3$) and a low initial entropy ($S^0<10~k_B/\mathrm{nucleon}$); in both cases, the r-process is weak due to the low neutron irradiation.

The relevance of the initial neutron mass fraction $X_n^0$ in the analysis of the efficiency of the r-process can be clearly seen in Figs.~\ref{fig:YAfSYe}.
The various trajectories can be classified according to their final abundance distributions (Figs.~\ref{fig:YAfSYe}a-f).
The trajectories computed within scenario~I are classified in the group labeled ``no $\nu$" while those computed within scenario~II are divided into five groups~: the group labeled ``fission" contains trajectories with the highest fission recycling ($X_\mathrm{fiss,tot}^\mathrm{cum}>0.01$, more details about the definition of this fission recycling indicator will be given in Sec.~\ref{sec:FR}).
Among the trajectories with $\sum_{125\leqslant A \leqslant 135} Y^\mathrm{f}(A)>0.01$, if $\sum_{190\leqslant A \leqslant 200} Y^\mathrm{f}(A)<0.001$ then they are classified in the group ``2\textsuperscript{nd} peak" else in the group ``rare earth \& no fission".
The trajectories with $\sum_{75\leqslant A \leqslant 85} Y^\mathrm{f}(A)>0.1$ are classified in the group ``1\textsuperscript{st} peak".
Remaining trajectories correspond to the group ``between 1\textsuperscript{st} and 2\textsuperscript{nd}peak".
The two first groups ``1\textsuperscript{st} peak" and ``between 1\textsuperscript{st} and 2\textsuperscript{nd}peak" represent each 7\% and 6\%, respectively, of the ejected mass within scenario~II (see Fig.~\ref{fig:YAfSYe}h).
The other groups ``2\textsuperscript{nd} peak", ``rare earth \& no fission" and ``fission" represent  41\%, 30\% and 16\%, respectively.
The mean initial neutron mass fraction increases with the group (Fig.~\ref{fig:YAfSYe}h, green curve), heavy elements production is possible only with a high initial neutron mass fraction. Similarly, the mean mass of elements heavier than 50 ($\left<A\right>_{A \geqslant 50}$) also increases with the group  (Fig.~\ref{fig:YAfSYe}h, gray curve) but reaches a maximum around 170 when fission occurs in groups ``fission" and ``no $\nu$" corresponding to the saturation regime induced by the fission recycling (see Sec.~\ref{sec:FR}).
In the ($S^0,Y_e^0$) plane (Fig.~\ref{fig:YAfSYe}g), the trajectories of the various groups are distributed according the iso-$X_n^0$ lines which clearly demonstrate the relevance of the initial neutron mass fraction $X_n^0$ in the analysis of the r-process efficiency.

\subsection{Fission recycling}
\label{sec:FR}

\begin{figure}
\includegraphics[width=\columnwidth]
{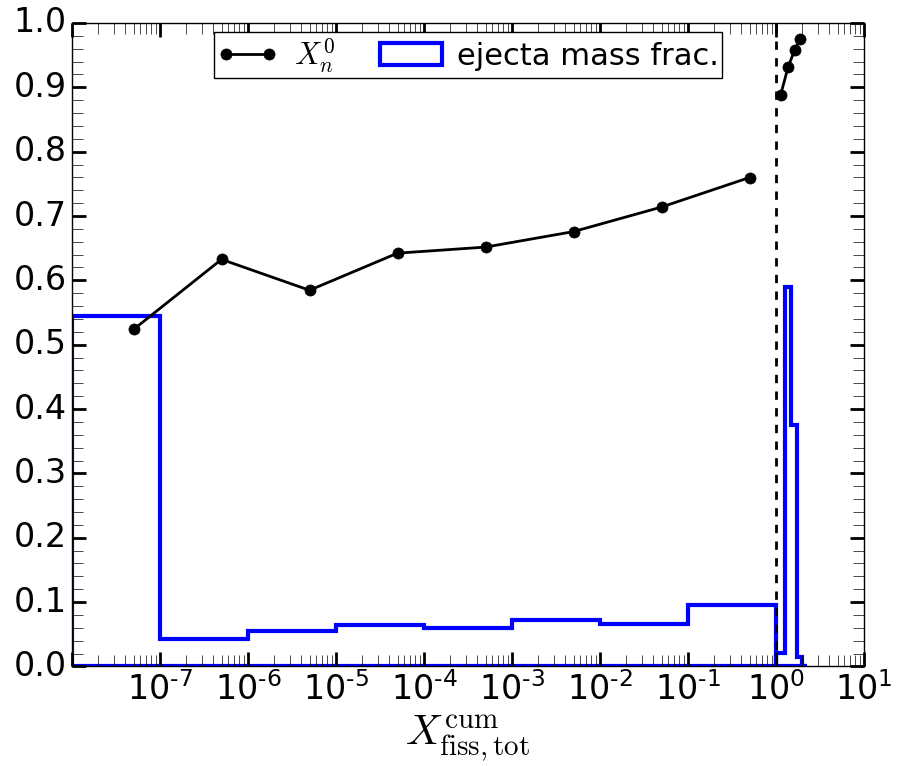}
\caption{(Color online) Initial neutron mass fraction $X_n^0$ (black curve with dots) and the mass distribution of the ejecta (blue curve) depending on the fission recycling $X_\mathrm{fiss,tot}^\mathrm{cum}$.
The dashed line separates abundances computed within scenario ~I ($X_\mathrm{fiss,tot}^\mathrm{cum}>1$) to the ones computed within scenario~II ($X_\mathrm{fiss,tot}^\mathrm{cum}<1$).}
\label{fig:Xn0_Xcumfiss}
\end{figure}

\begin{figure*}
\centering
	\subcaptionbox{For scenario~I where the mean value of the total fissioning cumulative mass fraction $\langle X_\mathrm{fiss,tot}^\mathrm{cum}\rangle_\mathrm{traj}=1.45$.\label{fig:YAfspynonu}}[\textwidth]
	{
	\includegraphics[width=\textwidth]
	{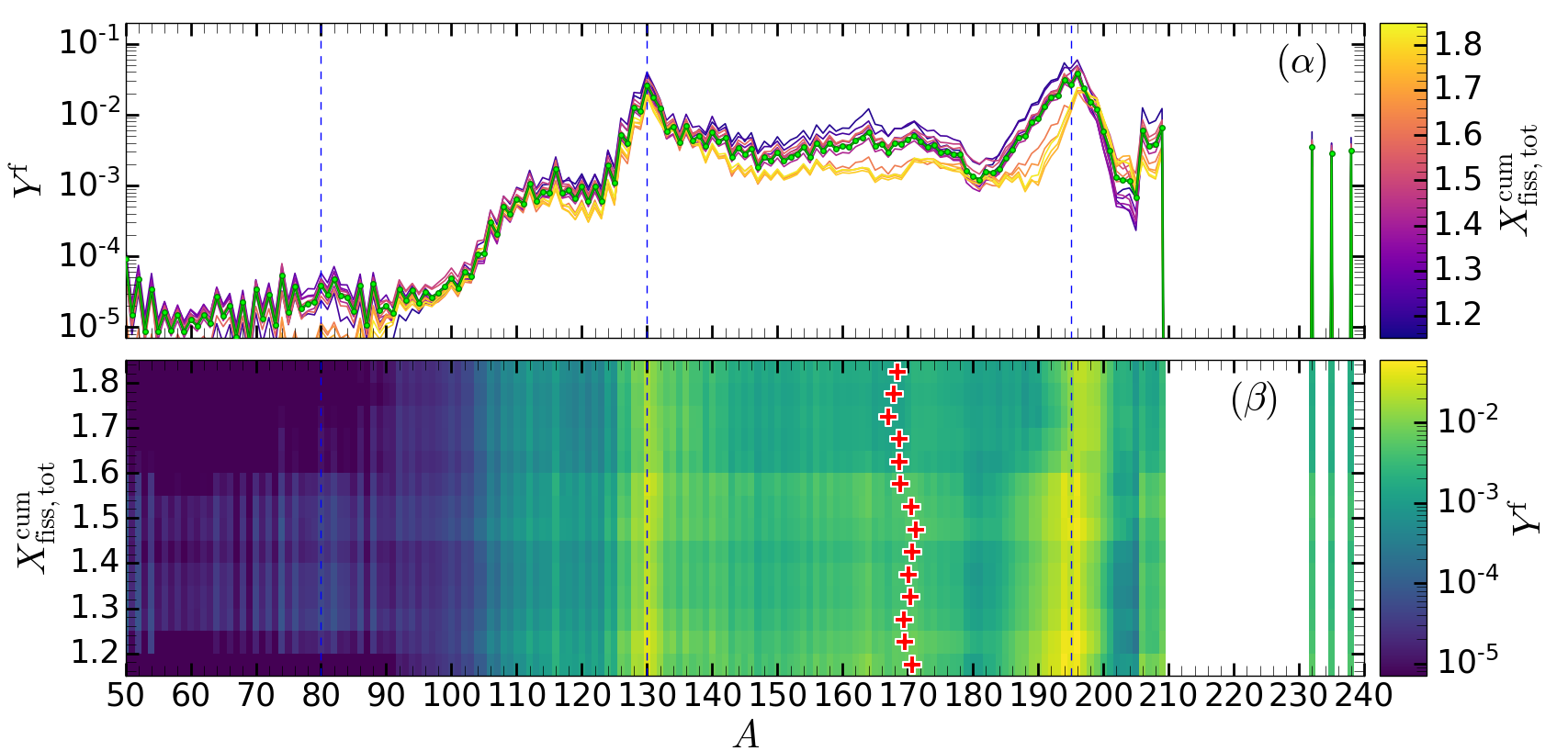}		
	}
\vfill \vspace{5mm}
	\subcaptionbox{For scenario~II where the mean value of the total fissioning cumulative mass fraction $\langle X_\mathrm{fiss,tot}^\mathrm{cum}\rangle_\mathrm{traj}=0.03$.\label{fig:YAfspyNUEC}}[\textwidth]
	{
	\includegraphics[width=\textwidth]
	{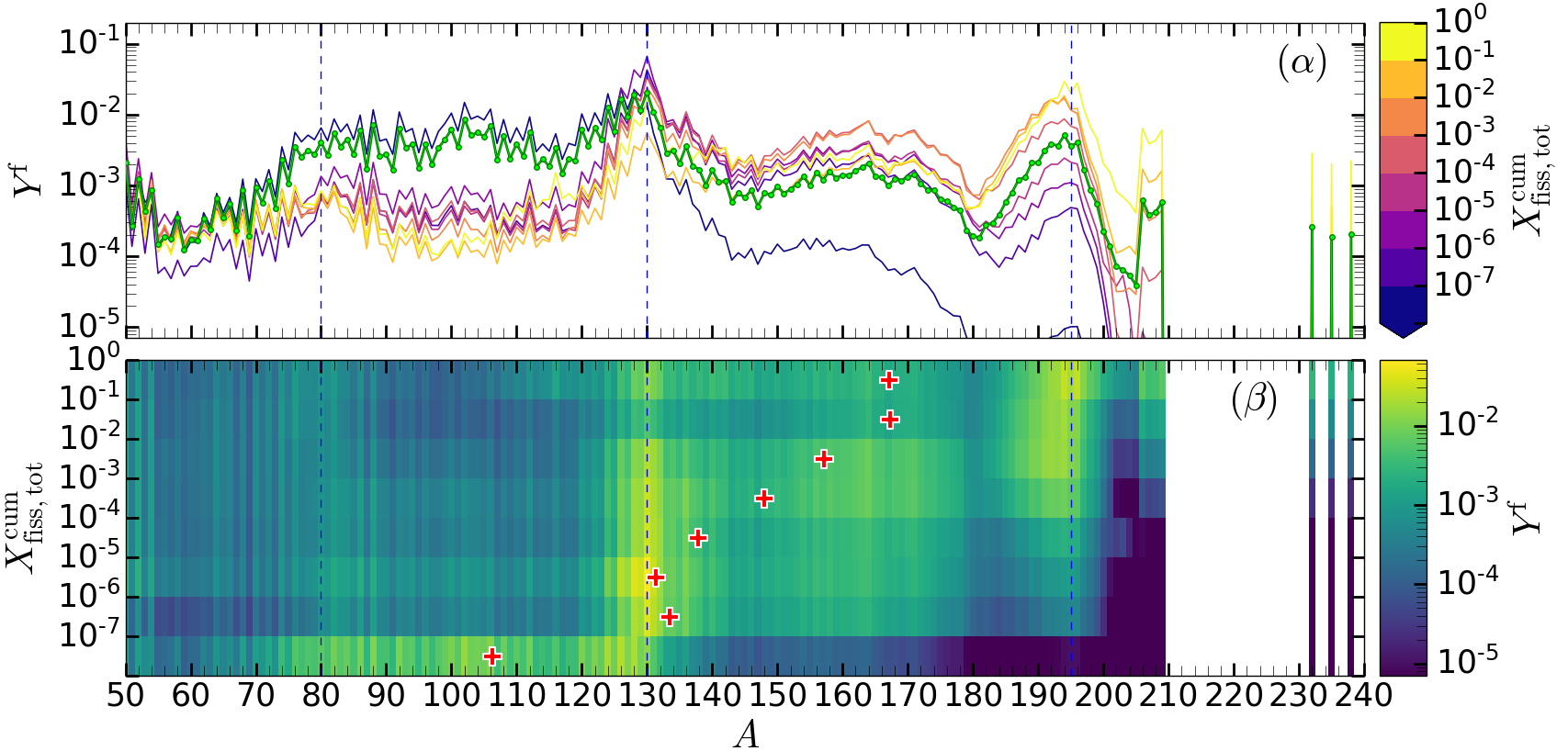}
	}
\caption{
(Color online) ($\alpha$) Final abundances obtained with SPY FFDs, averaged for each bin of $X_\mathrm{fiss,tot}^\mathrm{cum}$ colored by their $X_\mathrm{fiss,tot}^\mathrm{cum}$ value and compared to the average final abundances (green solid curve with dots). 
($\beta$) Same final abundances in the ($A,X_\mathrm{fiss,tot}^\mathrm{cum}$) plan colored by their abundances. 
The mean mass $\langle A\rangle_{A\geqslant 50}$ for each $X_\mathrm{fiss,tot}^\mathrm{cum}$ bins are represented by red crosses. 
}	
\label{fig:YAfspy}
\end{figure*}

Fission recycling for a given trajectory can be quantified by the so-called total cumulative fissioning mass fraction
\begin{eqnarray}
X_\mathrm{fiss,tot}^\mathrm{cum} &&= \sum_{Z,A} \sum_\mathrm{r=fiss} \int \frac{dX_C}{dt}(Z,A,t,r) dt \\
&& = \sum_{Z,A} \int  X(Z,A,t) \left[ N_n \langle \sigma v \rangle_\mathrm{n,f}^{Z,A} + \lambda_\mathrm{sf}^{Z,A}+ \lambda_{\beta \mathrm{df}}^{Z,A}\right] \; dt \nonumber \\
&& = \sum_{Z,A} X^\mathrm{cum}_\mathrm{fiss}(Z,A) \nonumber
\label{eq:defXfC}
\end{eqnarray}
where $\frac{dX_C}{dt}(Z,A,t,r)$ corresponds to the reaction flux, expressed in terms of the mass fraction $X$ for a given nucleus $(Z,A)$, at a given time $t$  and for a given fission reaction $r$. 
$X_\mathrm{fiss,tot}^\mathrm{cum}$ consequently corresponds, at a given time, to the cumulative mass fraction of fissioning nuclei that are destroyed by fission reactions and gives a quantitative indicator of the amount of mass recycled by fission processes.
Fission can take place by spontaneous  (at a rate $\lambda_\mathrm{sf}$) or $\beta$-delayed (at a rate $ \lambda_{\beta \mathrm{df}}$)  fission or by neutron-induced fission (at a rate $N_n \langle \sigma v \rangle_\mathrm{n,f}$, where $\langle \sigma v \rangle_\mathrm{n,f}$ is the corresponding astrophysical rate and $N_n$ the neutron density).
Consequently, there are three factors contributing to the fission recycling~: (\textit{i}) the quantity of fissionable nuclei, (\textit{ii}) the fission rates and (\textit{iii}) the amount of time over which fission reactions may occur.

The total fissioning cumulative mass fraction $X_\mathrm{fiss,tot}^\mathrm{cum}$ increases with initial neutron mass fraction $X_n^0$ (Fig.~\ref{fig:Xn0_Xcumfiss}, black curve with dots).
The fission recycling can therefore be explained in term of initial neutron mass fraction $X_n^0$, {\it i.e.} the more neutron-rich the initial ejecta, the larger the production of heavy elements, hence the more efficient the fission.

\subsubsection{Scenario~I}
\label{sec:FRsI}
Within scenario~I, $X_\mathrm{fiss,tot}^\mathrm{cum}>1$ for all ejected trajectories due to the high initial neutron mass fraction ($X_n^0>0.85$).
In Fig.~\ref{fig:YAfspynonu}, the final abundances of the 500 trajectories have been sorted and binned as a function of  $X_\mathrm{fiss,tot}^\mathrm{cum}$ which is found to range between 1.19 and 1.82 with a mean value over all trajectories of $\langle X_\mathrm{fiss,tot}^\mathrm{cum}\rangle_\mathrm{traj}=1.45$.
The final abundances distribution for a given trajectory can be directly linked to its total fissioning cumulative mass fraction $X_\mathrm{fiss,tot}^\mathrm{cum}$.
An overall shift of final abundances towards heavier elements can be expected with increasing fission recycling but the mean $A\geqslant 50$ mass, $\langle A\rangle_{A\geqslant 50}$, does not increase with $X_\mathrm{fiss,tot}^\mathrm{cum}$ (Fig.~\ref{fig:YAfspynonu}$\beta$, red crosses), and stays rather constant between 166 and 172. 
This value remains constant because the final abundances are dominated by the second ($A=130$) and third ($A=195$) peaks.

The final abundance of light elements ($A\leqslant 4$), mainly $\alpha$-particles, is significant with $0.1<\sum_{A\leqslant4}Y^\mathrm{f}(A)<0.84$ depending on the trajectory.
It increases with the initial entropy $S^0$, inducing an overall downward shift of the final abundances of heavier elements ($A>4$).
As $X_n^0$ is linked to $X_\mathrm{fiss,tot}^\mathrm{cum}$ (Fig.~\ref{fig:Xn0_Xcumfiss}), the initial entropy $S^0$ of trajectories within a given bin of $X_\mathrm{fiss,tot}^\mathrm{cum}$  covers almost the whole range of possible $S^0$ (since $X_n^0$ isolines evolve in the $(S^0,Y_e^0)$ plane, as seen in Fig.~\ref{fig:YAfSYe}g).
This $S^0$ admixture of trajectories within a bin of $X_\mathrm{fiss,tot}^\mathrm{cum}$ smooths the overall shift of the final abundances of heavy elements ($A>4$).
However, fission recycling has a significant impact~: the final $A<195$ abundances  decrease with $X_\mathrm{fiss,tot}^\mathrm{cum}$, whereas for elements $A>195$ they increase (Fig.~\ref{fig:YAfspynonu}$\alpha$).
With increasing values of  $X_\mathrm{fiss,tot}^\mathrm{cum}$, the abundances in the mass region $140<A<170$ drop by an order of magnitude, but around $A=202$ they increase from $4 \times 10^{-4}$ up to $2 \times 10^{-3}$, while  for $A<80$ elements, they drop below $10^{-5}$ (Fig.~\ref{fig:YAfspynonu}).

Neglecting the effect related to the variation of light elements abundance, a higher initial neutron mass fraction $X_n^0$ induces more fission recycling which implies a decrease of the final $A<80$ abundances.
A high neutron irradiation leads to a significant production of super-heavy elements ($A>300$) whose fission fragments combined to new neutron captures shift the third peak to higher masses.

Part of the nuclei that accumulated during neutron irradiation along the neutron drip line (including the neutron shell closure $N=184$) with $208<A \leqslant 278 $ are the progenitors of the trans-lead elements.
Among these accumulated nuclei, those with $260 \lesssim A \leqslant 278$ ($79 \lesssim Z \leqslant  94$) predominantly $\beta$-decay up to reach the ``decay roof 0" where they fission.
Those with $208<A<232$ $\beta$- and $\alpha$-decay up to the third abundance peak, while the remaining intermediate nuclei with $232 \leqslant A \lesssim 260$ $\beta$- and $\alpha$-decay up to the third abundance peak or $\beta$- decay and reach the ``decay roof 0" where they fission or decay into the long-lived ${}^{232}\mathrm{Th}$ and ${}^{235-238}\mathrm{U}$ nuclei.
Final $A>204$ abundances decrease with $X_\mathrm{fiss,tot}^\mathrm{cum}$ because their progenitors are the nuclei accumulated along the neutron shell closure $N=184$ during the neutron irradiation.
This accumulation is reduced for high neutron irradiations which enable to overcome the neutron shell closure $N=184$, hence to increase the fission recycling.

\subsubsection{Scenario~II}
\label{sec:FRsII}
With the low initial neutron mass fraction $X_n^0$ found in scenario~II, about $55\%$ of the ejected mass is not subject to fission recycling ($X_\mathrm{fiss,tot}^\mathrm{cum}<10^{-7}$) (Fig.~\ref{fig:Xn0_Xcumfiss}, blue curve) and the maximum value of $X_\mathrm{fiss,tot}^\mathrm{cum}=0.6$.
The fission recycling is reduced by a factor 50 with a mean value of the total fissioning cumulative mass fraction $\langle X_\mathrm{fiss,tot}^\mathrm{cum}\rangle_\mathrm{traj}$ that now amounts to 0.03 to be compared with 1.45 for scenario I.
The present scenario is well suited to study low fission recycling effects on the final abundances. 
For this reason, trajectories are sorted in $\log_{10}$ bins of their total fissioning cumulative mass fraction, as shown in Fig.~\ref{fig:YAfspyNUEC}.
The bins, except for the first one, represent each about $5\%$ to $10\%$ of the ejected mass (Fig.~\ref{fig:Xn0_Xcumfiss}).

The final abundance of the light $A\leqslant 4$ elements in bins are more or less similar ranging from 40\% to 55\% with no global trend which induces a smaller shift of the final abundances due to trans-alpha elements in comparison with scenario~I.

$X_\mathrm{fiss,tot}^\mathrm{cum}$ is linked to the initial neutron mass fraction $X_n^0$, as observed in Fig.~\ref{fig:Xn0_Xcumfiss} (black curve with dots) where $X_n^0$ is seen to increase linearly with $\log_{10}\left(X_\mathrm{fiss,tot}^\mathrm{cum}\right)$ from 0.5 to 0.75. In other words, $X_\mathrm{fiss,tot}^\mathrm{cum}$ increases exponentially with $X_n^0$.
An increase with $X_\mathrm{fiss,tot}^\mathrm{cum}$, hence with $X_n^0$, of the mean mass $\langle A\rangle_{A\geqslant 50}$ from 105 up to 168 is observed in Fig.~\ref{fig:YAfspyNUEC}$\beta$ (red crosses), {\it i.e.} the higher $X_n^0$, the heavier the final nuclei produced.

For trajectories with a low initial neutron mass fraction, $X_n^0\approx 0.5$ and $X_\mathrm{fiss,tot}^\mathrm{cum}<10^{-7}$, $A>130$ nuclei, including lanthanides, are not significantly produced in comparison with the other trajectories (Fig.~\ref{fig:YAfspyNUEC}) and production beyond $A>180$ is negligible. 
Fission recycling does not take place in this case because there are not enough neutrons to overcome significantly the $N=82$ neutron shell closure.
For higher initial neutron mass fraction $X_n^0\approx 0.6$ ($10^{-7}<X_\mathrm{fiss,tot}^\mathrm{cum}<10^{-5}$), nuclei in the rare earth and third peak regions start to be produced, but those with $A>200$ are still highly underproduced and fission recycling remains negligible.

Abundances of third peak elements increase in the range $10^{-5}<X_\mathrm{fiss,tot}^\mathrm{cum}<10^{-1}$ ($0.65<X_n^0<0 .71$) as does the mean mass $\langle A\rangle_{A\geqslant 50}$ which increases linearly with $\log_{10}\left(X_\mathrm{fiss,tot}^\mathrm{cum}\right)$ up to 168.
In this $X_\mathrm{fiss,tot}^\mathrm{cum}$ range, abundances of rare earth nuclei ($145<A<180$) increase to reach a maximum at $10^{-4}<X_\mathrm{fiss,tot}^\mathrm{cum}<10^{-2}$ before decreasing for $X_\mathrm{fiss,tot}^\mathrm{cum}>10^{-2}$.
The $N=126$ and $N=184$ neutron shell closures start to be overcome for $X_\mathrm{fiss,tot}^\mathrm{cum}>10^{-4}$ leading to the synthesis of $A>204$ nuclei in a non-negligible amount. 
Fission recycling starts to play a role for trajectories with $X_\mathrm{fiss,tot}^\mathrm{cum}>10^{-4}$ (or equivalently with $X_n^0>0.65$), which represent $29\%$ of the ejected mass in scenario~II.

In the last bin of total fissioning cumulative mass fraction, {\it i.e.} $0.1<X_\mathrm{fiss,tot}^\mathrm{cum}<0.61$ and an average initial neutron mass fraction $X_n^0=0.76$, the third r-abundance peak becomes higher than the second one and $A>204$ nuclei dominate the composition.
The mean mass $\langle A\rangle_{A\geqslant 50}$ does not increase significantly, a saturation regime is reached with $\langle A\rangle_{A\geqslant 50}\approx 170$ as observed for scenario~I (see Sec.~\ref{sec:FRsI}); the r-process dynamics for such trajectories is similar to the one taking place in scenario~I.

To complete the discussion of Sec.~\ref{sec:FRsI}, fission recycling plays a dominant role when $X_\mathrm{fiss,tot}^\mathrm{cum}>0.1$ corresponding to $X_n^0>0.75$ which concerns $9.5\%$ of the ejected mass within scenario~II.
A saturation regime is reached $\langle A\rangle_{A\geqslant 50}\approx 170$, when a high initial neutron mass fraction $X_n^0$ induces more fission processes feeding back the $80<A<200$ mass region.

\subsection{Fission seeds}
\label{sec:FS}

\begin{figure}
\includegraphics[width=\columnwidth]
{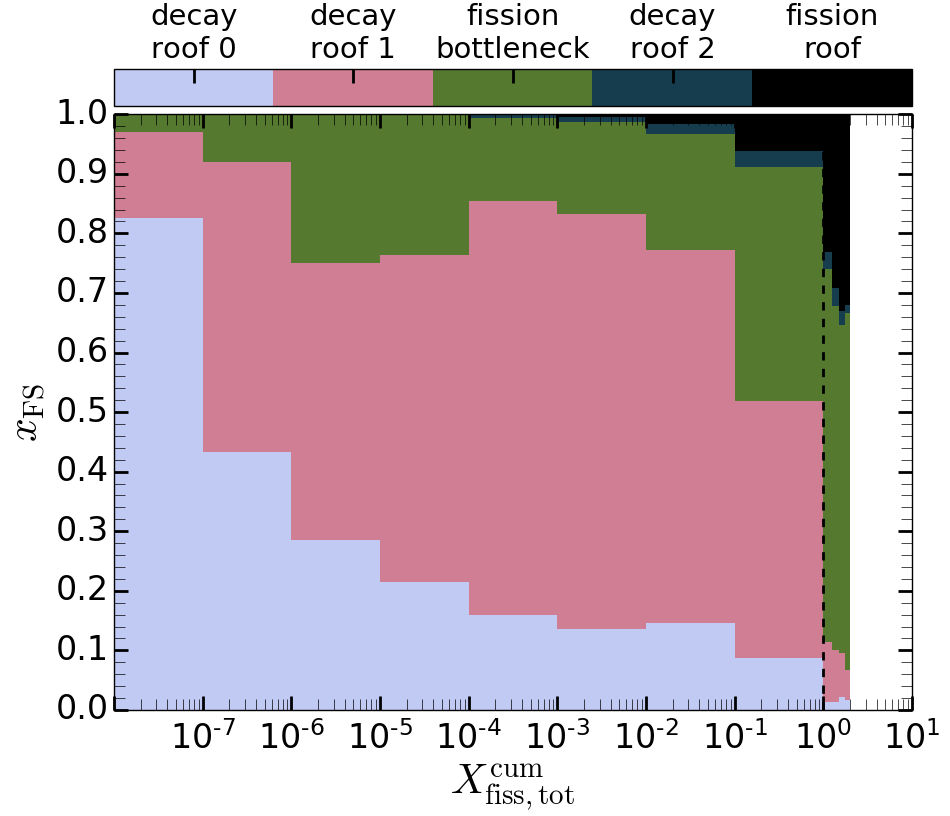}
\caption{
(Color online) Repartition of the fission recycling $X_\mathrm{fiss,tot}^\mathrm{cum}$ between the 5 fission seed areas defined in Fig.~\ref{fig:bknbdfsfa}d.
The dashed line separates scenario~I ($X_\mathrm{fiss,tot}^\mathrm{cum}>1$) from scenario~II ($X_\mathrm{fiss,tot}^\mathrm{cum}<1$).
}
\label{fig:fisslim_Xcumfiss}
\end{figure}

\begin{figure*}
\centering
	\subcaptionbox{For scenario~I.\label{fig:FissSeeds_SPY_noNU}}
	{	
	\includegraphics[width=0.9\textwidth]
	{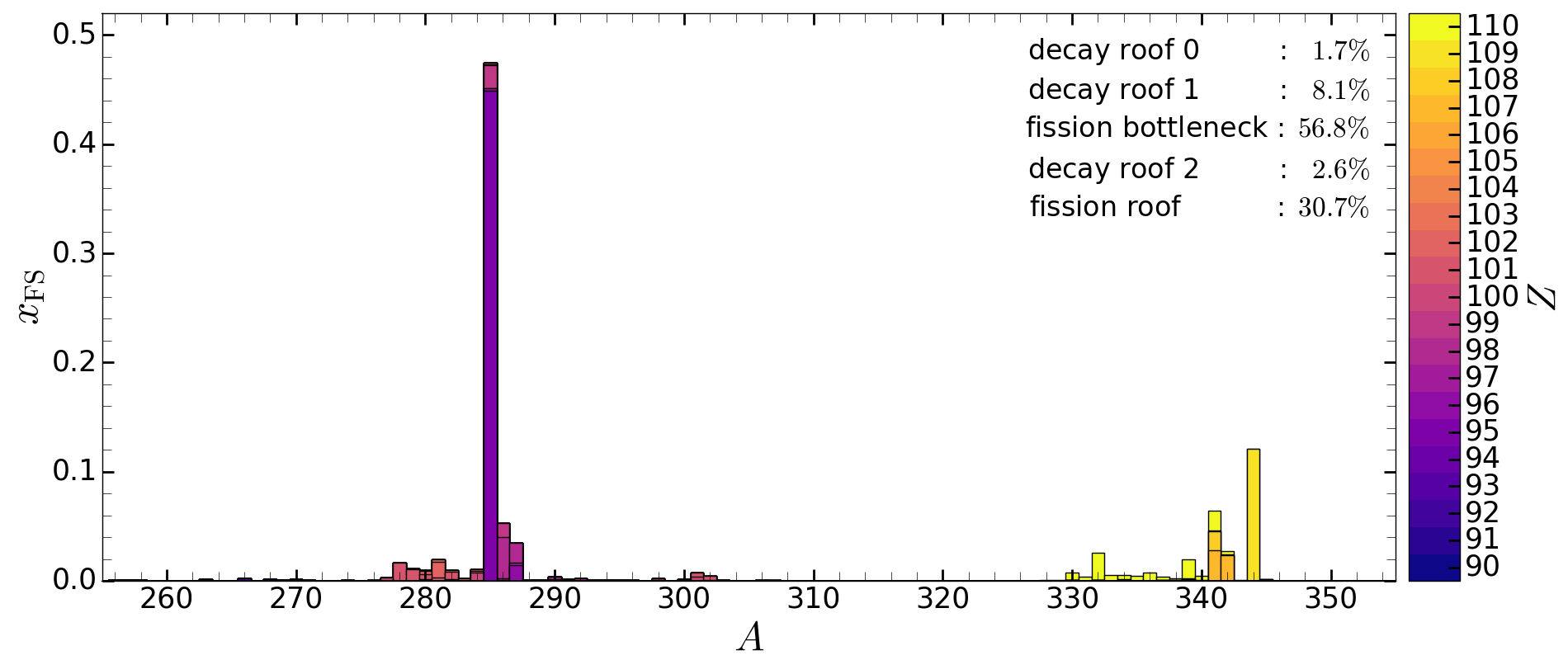}	
	}
	\vfill\vspace{5mm}
	\subcaptionbox{For scenario~II.\label{fig:FissSeeds_SPY_NuEC}}
	{	
	\includegraphics[width=0.9\textwidth]
	{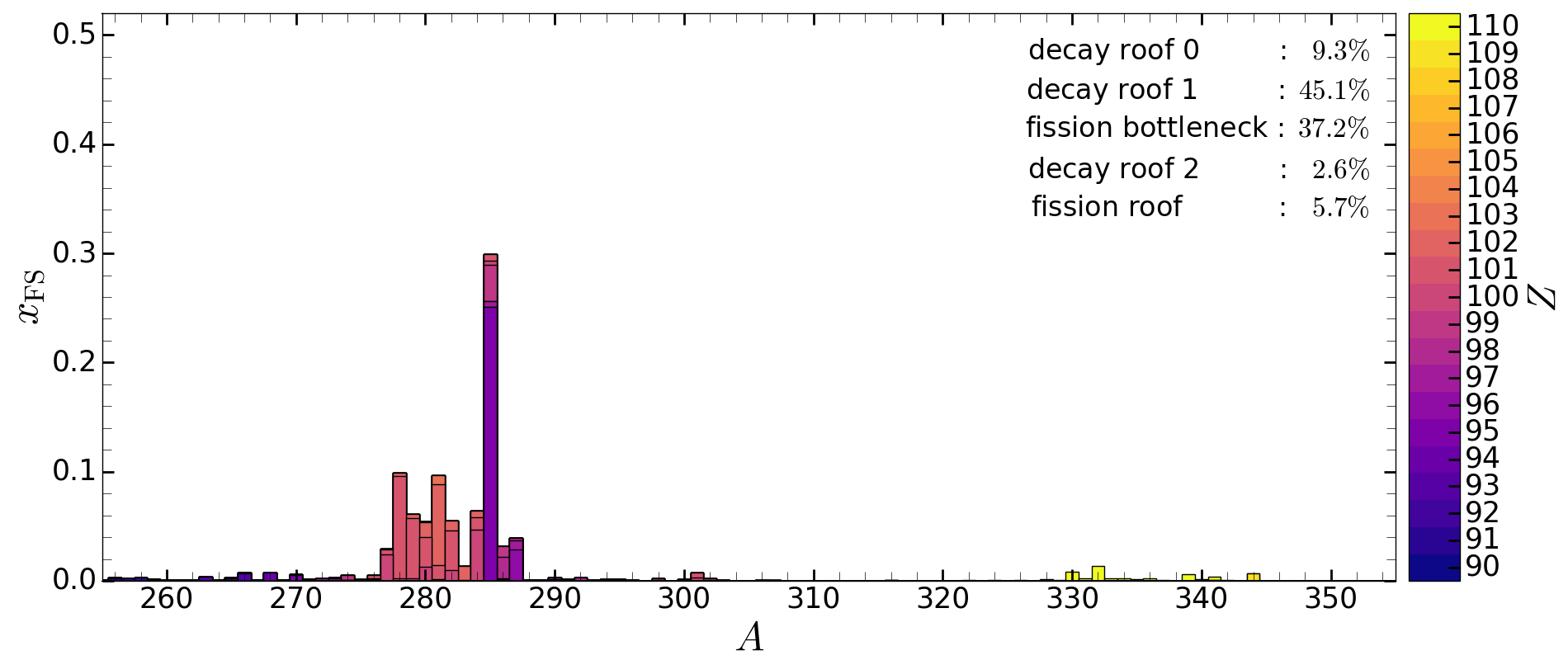}
	}
\label{fig:FissSeeds_SPY}
\caption{
(Color online) Fission seed contribution ($x_\mathrm{FS}$) depending on the mass of fission seed (the total height). 
Each bins contains the isotopic distribution of $x_\mathrm{FS}$ for a given A.
}
\end{figure*}

We now turn to the mean contribution of fissioning nuclei, or fission seeds, to the fission recycling and to the final abundance distribution.
The contribution of a given fissioning nucleus $(Z,A)$ to the fission recycling, $x_\mathrm{FS}(Z,A)$, can be defined, for a given set of trajectories, as the fraction of the total fissioning cumulative mass fraction $\langle X_\mathrm{fiss,tot}^\mathrm{cum}\rangle_\mathrm{traj}$ due to this specific fissioning nucleus ($Z,A$), {\it i.e.}
\begin{equation}
x_\mathrm{FS}(Z,A)=\frac{\langle X^\mathrm{cum}_\mathrm{fiss.}(Z,A) \rangle_\mathrm{traj}}{\langle X_\mathrm{fiss,tot}^\mathrm{cum}\rangle_\mathrm{traj}} ~.
\label{eq:x_FS}
\end{equation}
In the specific NS merger cases studied here, $\langle X_\mathrm{fiss,tot}^\mathrm{cum}\rangle_\mathrm{traj}$ reaches rather similar values of 1.45 and 1.41 in scenario~I when using SPY or GEF FFDs, respectively and 0.03 for both FFDs within scenario~II.
This similitude is due to the fact that SPY and GEF FFDs are not fundamentally different (see Fig.~\ref{fig:YAspygef}) and the properties of the fission seeds are not related to their FFDs (see Fig.~\ref{fig:bknbdfsfa}). Consequently, only results with SPY FFDs are presented here.

\subsubsection{Location in the $(N,Z)$ plane of the contributing fission seeds}
\label{sec:FSfsa}
Due to the specific  $\beta$-decay and neutron capture properties of nuclei above Th, in particular the ``fission bottleneck", the ``fission roof" and the $N=184$ neutron shell closure, not all the trans-Th nuclei play a role in fission recycling.
The contribution of a given fission seed area  (see Fig.~\ref{fig:bknbdfsfa}d defining the 5 areas of relevance for the nuclear physics adopted in the present study) to the fission recycling $X_\mathrm{fiss,tot}^\mathrm{cum}$ is the sum of $x_\mathrm{FS}(Z,A)$ for all nuclei located in such a specific fission seed area. 
Such contributions evolve with the fission recycling, as shown in Fig.~\ref{fig:fisslim_Xcumfiss}.
The contribution of the ``decay roof 0", corresponding to the fission of nuclei originally accumulated along the neutron shell closure $N=184$ and $Z<94$ and reaching the fission efficient region by $\beta$-decay after the freeze-out, decreases with increasing $X_\mathrm{fiss,tot}^\mathrm{cum}$ since the fission recycling increases with the initial neutron mass fraction $X_n^0$.
With increasing $X_\mathrm{fiss,tot}^\mathrm{cum}$ and $X_n^0$, matter accumulated along $N=184$ isotone spreads up to Fm thanks to the successive $\beta$-decays and neutron captures along this shell closure.
More nuclei above $Z=94$ are produced and increase the contribution of the ``decay roof 1" and consequently decrease the one of the ``decay roof 0".
For $X_\mathrm{fiss,tot}^\mathrm{cum}>10^{-6}$, the neutron shell closure $N=184$ starts to be overcome and nuclei in the ``fission bottleneck" area contribute to the fission recycling.
In the range $10^{-6}<X_\mathrm{fiss,tot}^\mathrm{cum}<10^{-1}$, the contribution of the ``fission bottleneck" is roughly constant representing 20\% of the total cumulative fissioning mass fraction.
The matter starts to flow through the ``fission bottleneck", via Cm isotopes, and reach the ``fission roof" for $X_\mathrm{fiss,tot}^\mathrm{cum}>10^{-4}$.
For high fission recycling ($X_\mathrm{fiss,tot}^\mathrm{cum}>1$ as in scenario~I), the total cumulative fissioning mass fraction is dominated by the ``fission bottleneck" and  ``fission roof" areas, representing around 60\% and 30\%, respectively.
The contribution of the ``decay roof 2" is low because matter does not pile up in this region due to the absence of shell closure.

\subsubsection{Scenario~I}
\label{sec:FSsI}

Sixteen fission seeds are found to contribute for more than $1\%$ to the fission recycling (Fig.~\ref{fig:FissSeeds_SPY_noNU}) within scenario~I. They are distributed  mainly in the ``fission bottleneck" and the ``fission roof" areas, representing respectively 57\% and 31\% of the total cumulative fissioning mass fraction (Fig.~\ref{fig:FissSeeds_SPY_noNU}).

A major part of the flow crossing the $N=184$ shell closure is concentrated along the Am chain, fed by the favorable $\beta$-decay of ${}^{278}_{94}\mathrm{Pu}_{184}$, where ${}^{285}_{95}\mathrm{Am}_{190}$ with $x_\mathrm{FS}=45\%$ is the major contributor to the fission recycling. 
This favorable fission is due to the low fission barriers of a few nuclei dropping below 4~MeV according to the BSk14 Skyrme HFB calculations \cite{Goriely09b}.
They are also main contributors in the ``fission roof" area, the main one in this area and the second major one of all the contributors is ${}^{344}_{109}\mathrm{Mt}_{233}$ with $x_\mathrm{FS}=12\%$. In this case, the primary fission barrier drop below 2~MeV.
As the main fission seeds are in the $275<A<290$ and $330<A<346$ ranges, the FFDs of these fission seeds affect the final production of nuclei in the $80<A<210$ range.

\subsubsection{Scenario~II}
\label{sec:FSsII}

As the initial neutron mass fraction is low in this case, the fission recycling remains low too.
The contribution of the ``fission bottleneck" and the  ``fission roof" amount to 37\% and 6\%, respectively, to be compared with 57\% and 31\% in scenario~I.
The main fission seeds area is the ``decay roof 1" (Fig.~\ref{fig:FissSeeds_SPY_NuEC}) because the low neutron irradiation makes it difficult to overcome the $N=184$ neutron shell closure.
It represents 45\% of $X_\mathrm{fiss,tot}^\mathrm{cum}$ and the main contributors are Md and No isotopes.
The contribution of the ``decay roof 0" increases from $2\%$ in scenario~I to the non-negligible value of $9\%$ in scenario~II.

The main fission seed is still ${}^{285}_{95}\mathrm{Am}_{190}$ with $x_\mathrm{FS}=25\%$ with additional ones located in the $276<A_\mathrm{CN}<288$ mass region with $x_\mathrm{FS}=2-10\%$.
The fission recycling occurs mostly after freeze-out in scenario~II.
The post-freeze-out fission seeds, located in the ``decay roof 0 and 1", represent about 54\% of $X_\mathrm{fiss,tot}^\mathrm{cum}$.

\subsection{Fission progenitors}
\label{sec:FP}

\begin{figure*}
\includegraphics[width=0.8\textwidth]
{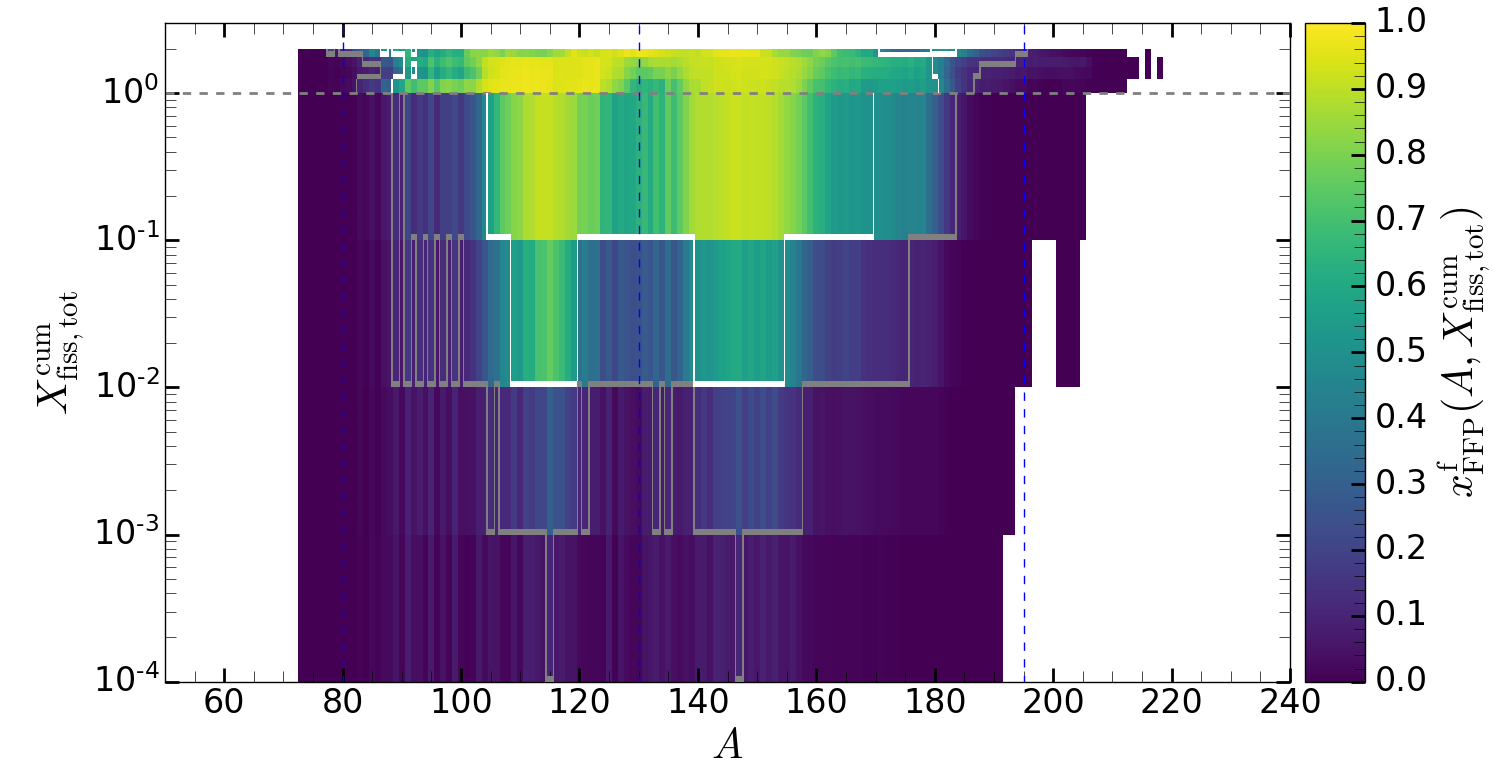}
\caption{
(Color online) Contribution of the fission fragments to the final production of nuclei with mass number $A$ $x_\mathrm{FFP}^\mathrm{f}(A,X_\mathrm{fiss,tot}^\mathrm{cum})$ obtained with SPY FFDs.
Gray curve delimits $x_\mathrm{FFP}^\mathrm{f}(A,X_\mathrm{fiss,tot}^\mathrm{cum})=0.1$ and white one $x_\mathrm{FFP}^\mathrm{f}(A,X_\mathrm{fiss,tot}^\mathrm{cum})=0.5$.
}
\label{fig:fffp_Xcumfiss}
\end{figure*}

We now turn in the analysis of the so-called fission progenitors, {\it i.e.} the nuclei for which fission contributes in a non-negligible way to their final abundance.
Through a detailed tracking of the fission processes taking place during the r-process, the impact of fission on the final abundances can be quantified by the analysis of the fission fragments progenitors and their related fission seeds.
For each trajectory, the progenitors of a given fission fragment can be identified and the fission contribution to the production of this final nucleus can be estimated.
More precisely, the contribution stemming from the fission fragments $x_\mathrm{FFP}(Z,A,t_i)$ (FFP stands for fission fragments progenitors) to a nucleus $(Z,A)$ with the molality $Y_m(Z,A,t_i)$ (in mol/g) at time $t_i$ is defined as
\begin{widetext}
\begin{equation}
x_\mathrm{FFP}(Z,A,t_i)=\frac{
x_\mathrm{FFP}(Z,A,t_{i-1}) Y_m(Z,A,t_{i-1})+\frac{\Delta Y_{m,\mathrm{P}}}{\Delta t}(Z,A,t_{i-1})\Delta t\sum_{r}x_\mathrm{FFP}(Z_r,A_r,t_{i-1})x_\mathrm{RP}(Z,A,r,t_{i-1})
}{Y_m(Z,A,t_{i-1})+\frac{\Delta Y_{m,\mathrm{P}}}{\Delta t}(Z,A,t_{i-1})\Delta t}
\label{eq:fisstracking}
\end{equation}
\end{widetext}
where $\Delta t=t_i-t_{i-1}$ is an arbitrary time step taken here to be of the order of  $\approx t_i/10$ and $\frac{\Delta Y_{m,\mathrm{P}}}{\Delta t}(Z,A,t_{i-1})$ is the total production rate (in mol/g/s) of nucleus $(Z,A)$ between $t_{i-1}$ and $t_i$, expressed in terms of the molality $Y_m$ (if normalized to 1, it corresponds to the molar fraction $Y$).
The reaction fraction $x_\mathrm{RP}(Z,A,r,t_{i-1})$ is the fraction of the total production rate due to the reaction $(r)$.
The  reactions considered here are  $(n,\gamma)$, $(\gamma,n)$, $(\beta,\gamma)$, $(\beta,n)$, $(\beta,2n)$, $(\beta,3n)$, $(\gamma,\alpha)$, (sf) ({\it i.e.} spontaneous fission), $(\beta,\mathrm{f})$ and $(n,\mathrm{f})$.
In other words, the contribution of fission fragments to the production of a nucleus ${}_{Z}^{A}\mathrm{X}$ at time $t_i$ is the weighted average between the previous contribution $x_\mathrm{FFP}(Z,A,t_{i-1})$ and the contributions $x_\mathrm{FFP}(Z_r,A_r,t_{i-1})$ due to the production of this element by a given reaction $r~: {}_{Z_r}^{A_r}\mathrm{Y} \rightarrow {}_{Z}^{A}\mathrm{X}+...$, between $t_{i-1}$ and $t_i$, at a rate of $\frac{\Delta Y_{m,\mathrm{P}}}{\Delta t}(Z,A,t_{i-1})x_\mathrm{RP}(Z,A,r,t_{i-1})$.
If a nucleus is exclusively produced by fission, $x_\mathrm{FFP}=1$.

This quantity is well suited to quantify the impact of fission fragments on final abundances because it is not affected by the decay of post-neutron fission fragments thanks to the time tracking.
To obtain the overall final $x_\mathrm{FFP}$, hereafter $x_\mathrm{FFP}^\mathrm{f}$, (Figs.~\ref{fig:FFD_SPYvsGEF}, red dotted curves), the tracking (Eq.~\ref{eq:fisstracking}) is performed up to $t=1$~yr when the residual fission does not contribute anymore to the production of light nuclei.

The contribution of fission fragments to the final production depends on the fission recycling indicator $X_\mathrm{fiss,tot}^\mathrm{cum}$ (Fig.~\ref{fig:fffp_Xcumfiss}, gray curve) and starts to be significant, \textit{i.e.}
\begin{equation}
\max_{A}\left[x_\mathrm{FFP}^\mathrm{f}(A,X_\mathrm{fiss,tot}^\mathrm{cum})\right](X_\mathrm{fiss,tot}^\mathrm{cum})>0.1, \nonumber
\end{equation}
for $X_\mathrm{fiss,tot}^\mathrm{cum}>10^{-4}$ ($X_n^0>0.65$) which corresponds to the value given in Sec.~\ref{sec:FRsII}.
For $X_\mathrm{fiss,tot}^\mathrm{cum}>10^{-2}$, fission contributes at least to 50\% of the production of the final nuclei with $108<A<120$ and $139<A<155$.
For a high fission recycling ($X_\mathrm{fiss,tot}^\mathrm{cum}\approx 1$), this contribution can reach nearly 100\% for the elements with $A\approx 110$.

In addition to the strong sensitivity to the fission recycling, $x_\mathrm{FFP}^\mathrm{f}(A)$ is obviously affected by the FFDs model adopted, even if the fission recycling does not depend significantly on the FFDs model.
This sensitivity to the FFDs model is studied for both scenario in the two subsections below.
 In order to clarify the link between $x_\mathrm{FS}$ and $x_\mathrm{FFP}^\mathrm{f}$, we also introduce here two new quantities, namely the relative contribution of each fission seed to the production of a final nucleus $A$ by fission  
\begin{equation}
x^\mathrm{f}_\mathrm{FS/FFP}(A,A_\mathrm{FS})=\frac{x_\mathrm{FFP}^\mathrm{f}(A,A_\mathrm{FS})}{x_\mathrm{FFP}^\mathrm{f}(A)}
\label{eq:x_FS2FFP}
\end{equation}
where $x_\mathrm{FFP}^\mathrm{f}(A,A_\mathrm{FS})$ is obtained by tracking only fission fragments produced by the specific fission  seed $A_\mathrm{FS}$, $x_\mathrm{RP}(Z,A,t_{i-1})$ being replaced by $x_\mathrm{RP}(Z,A,A_\mathrm{FS},t_{i-1})$ in Eq.~\ref{eq:fisstracking}.
and the contribution of a compound nucleus $(Z_\mathrm{CN},A_\mathrm{CN})$ to the fission recycling, which can be expressed from Eq.~\ref{eq:x_FS} as 
\begin{equation}
x_\mathrm{CN}(Z_\mathrm{CN},A_\mathrm{CN})=\sum_{(Z,A)_\mathrm{CN},r=\left\lbrace\mathrm{sf,\beta df,nif}\right\rbrace} x_\mathrm{FS}(Z_r,A_r,r)
\label{eq:x_CN}
\end{equation}
where $x_\mathrm{FS}(Z_r,A_r,r)$ is the contribution to the fission recycling $\langle X_\mathrm{fiss,tot}^\mathrm{cum}\rangle_\mathrm{traj}$ of the fission seed $(Z_r,A_r)$ fissioning by a given fission mode  $r$ (spontaneous, $\beta$-delayed or neutron induced).
In Eq.~\ref{eq:x_CN}, the summation is performed in such a way that the mass of the compound nucleus formed from the fission seed is equal to $A_\mathrm{CN}$ and the proton number to $Z_\mathrm{CN}$.

\begin{figure*}
\centering
  \subcaptionbox{Around the neutron shell closure $N=184$, these nuclei are responsible for 64/86\% of the fission recycling in scenario~I/II.
  \label{fig:spygef_93104}}
  {  
  \includegraphics[width=\textwidth]{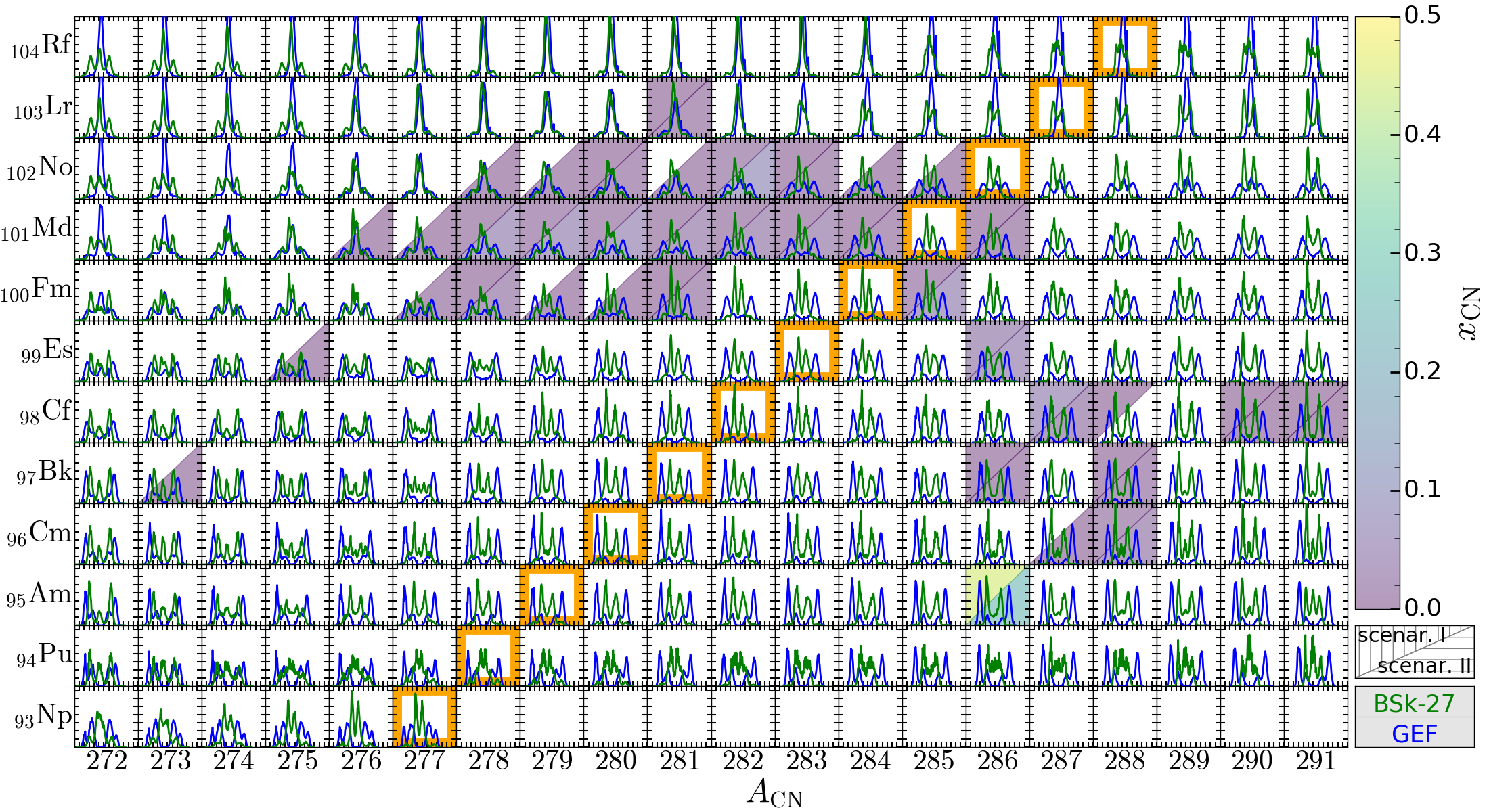}
  }  
  \vfill
  \vspace{5mm}
  \subcaptionbox{Around the ``fission roof", these nuclei are responsible for 30/5\% of the fission recycling in scenario~I/II.
  \label{fig:spygef_99110}}
  {  
  \includegraphics[width=\textwidth]{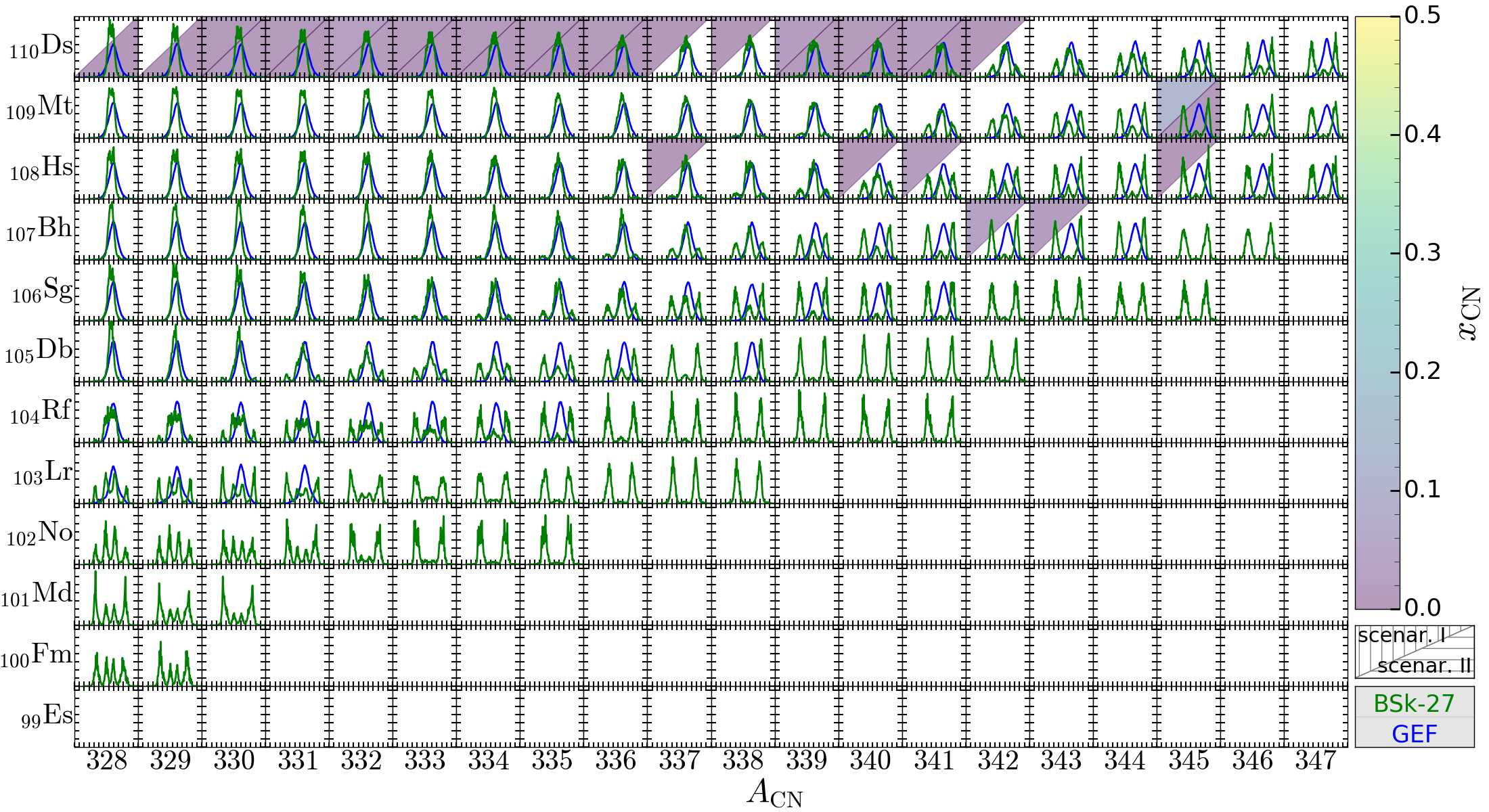}
  }    
  \caption{(Color online) Post-neutron FFDs from SPY model computed at a constant excitation energy $Q=8$~MeV (green curves) and from GEF model for thermal neutron-induced fission (blue curves).  
  For each box, the x axis ranges from $A=80$ to 210 by step of 10 and the y axis from 0 to 15\% by step of 2\%. 
  Colored background correspond to the contribution of the compound nucleus $x_\mathrm{CN}$ in scenario~I/II.
  Only contributions $x_\mathrm{CN}>0.001$ are displayed.
  FFDs in thick orange boxes correspond to the $N_\mathrm{CN}=184$ neutron shell closure.}
  \label{fig:YAspygef}
\end{figure*}

\subsubsection{Scenario~I}
\label{sec:FPsI}

\begin{figure*}
\centering
	\subcaptionbox{For scenario~I.\label{fig:FFD_SPYvsGEF_noNU}}
	{	
	\includegraphics[width=\columnwidth]{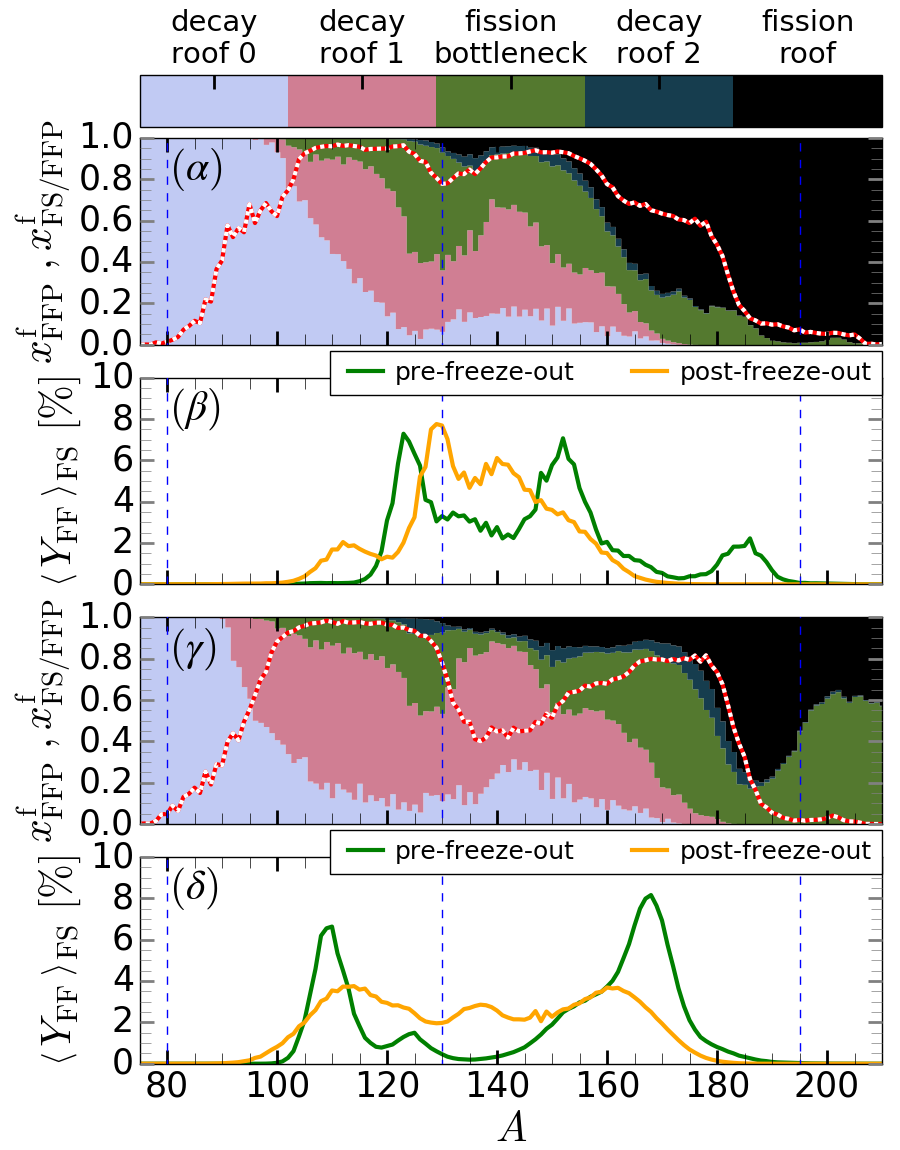}
	}
	\hfill	
	\subcaptionbox{For scenario~II.\label{fig:FFD_SPYvsGEF_NuEC}}
	{	
	\includegraphics[width=\columnwidth]{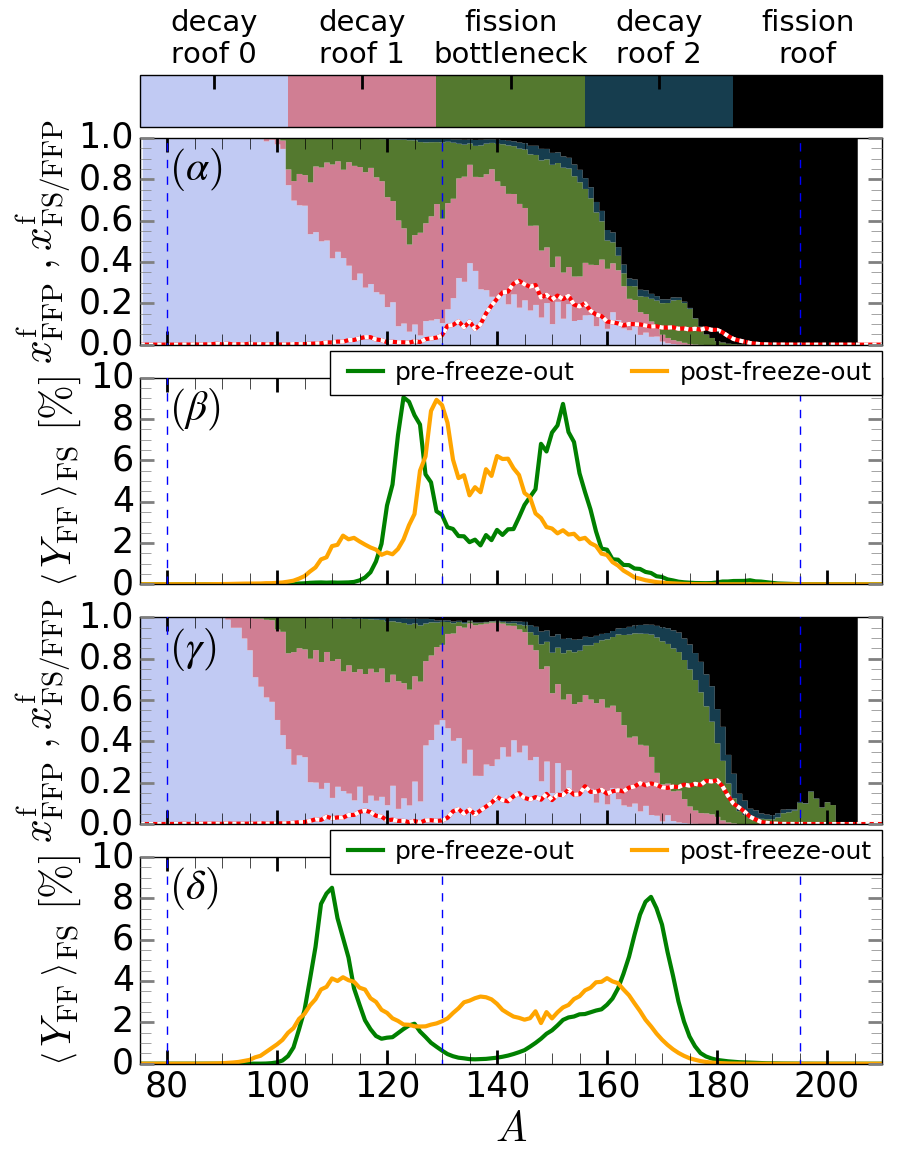}
	}		
\caption{
(Color online) ($\alpha$ and $\gamma$) Contribution of the fission fragments to the final production ($x^\mathrm{f}_\mathrm{FFP}$) (red dotted curves) and  fission seeds contributions to the production of fission fragment progenitors ($x^\mathrm{f}_\mathrm{FS/FFP}$) (color scale) computed with SPY($\alpha$) or GEF ($\gamma$) FFDs .
($\beta$ and $\delta$) Mean fission yields $\left< Y_\mathrm{FF}\right>_\mathrm{FS}$ for SPY  ($\beta$) or GEF model ($\delta$) averaged over fission seeds from the ``fission bottleneck" and the ``fission roof" (green curves, denoted as pre-freeze-out) or from ``decay roofs" (orange curves, denoted as post-freeze-out).
}
\label{fig:FFD_SPYvsGEF}
\end{figure*}

In scenario~I,  fission processes contribute significantly to the production of nuclei with $80<A<200$ for both FFDs models (Fig.~\ref{fig:FFD_SPYvsGEF_noNU}, red dotted curves) due to the high fission recycling.
The shape of $x^\mathrm{f}_\mathrm{FFP}(A)$ is affected by the FFDs model (Figs.~\ref{fig:FFD_SPYvsGEF_noNU}$\alpha$ and $\gamma$, red dotted curves).
However the link between $x^\mathrm{f}_\mathrm{FFP}$ and FFDs is no straightforward.

Two mean fission yields distributions $\left<Y_\mathrm{FF}\right>_\mathrm{FS}$ can be defined, the first one is associated with the ``fission bottleneck" and the ``fission roof", denoted as pre-freeze-out mean fission yields, corresponds to the mean fission yields distribution during the neutron irradiation, before  freeze-out (Figs.~\ref{fig:FFD_SPYvsGEF}, green curves).
The second one, denoted as post-freeze-out mean fission yields, is associated with the ``decay roofs" (Figs.~\ref{fig:FFD_SPYvsGEF}, orange curves), after freeze-out.

For both FFDs model, the pre-freeze-out mean fission yields is dominated by the FFDs of \textsuperscript{286}Am ($x_\mathrm{CN}=0.45$) and \textsuperscript{345}Mt ($x_\mathrm{CN}=0.12$) formed by the neutron capture of \textsuperscript{285}Am and \textsuperscript{344}Mt.
With SPY FFDs (see Figs.~\ref{fig:YAspygef}, green curves), these two main contributors fission asymmetrically. The three peaks pattern seen in the pre-freeze-out $\left<Y_\mathrm{FF}\right>_\mathrm{FS}$  (Fig.~\ref{fig:FFD_SPYvsGEF_noNU}$\beta$) is due to (\textit{i}) the light fragments of \textsuperscript{286}Am feeding the peak around $A=120-130$, (\textit{ii}) the heavy fragments of \textsuperscript{286}Am plus the light fragments of \textsuperscript{345}Mt feeding the peak around $A=145-160$, and (\textit{iii}) the heavy fragments of \textsuperscript{346}Mt responsible for the peak $A=180-190$.
In contrast, the GEF model (see Figs.~\ref{fig:YAspygef}, blue curves) predicts a more asymmetric fission of \textsuperscript{286}Am but a more symmetric one for \textsuperscript{346}Mt.
As a result, a double hump pattern is obtained for the pre-freeze-out $\left<Y_\mathrm{FF}\right>_\mathrm{FS}$ (Fig.~\ref{fig:FFD_SPYvsGEF_noNU}$\delta$) where the peak around $A=100-120$ is due to to the light fragments of \textsuperscript{286}Am and the one around $A=150-180$ to the heavy fragments of \textsuperscript{286}Am plus the symmetric fragments of \textsuperscript{346}Mt.
 
The main contributors to the post-freeze-out mean fission yields, {it i.e.} to the fission recycling in the ``decay roofs" areas, are isotopes from Fm to Lr with $A_\mathrm{CN}=278$ to 284 where $x_\mathrm{CN}$ reaches values up to 0.016 for \textsuperscript{283}No (see Fig.~\ref{fig:spygef_93104}).
Those nuclei are characterized by an  admixture of slightly asymmetric fission and a triple-peak fission where the maximum yields are found around $A=130$ for SPY FFDs. GEF FFD is more asymmetric leading to post-freeze-out mean fission yields roughly constant in the $A=110-160$ mass range.

During the neutron irradiation, fission fragments captures neutrons.
The $A<130$ fission fragments  accumulate along the neutron shell closure $N=82$ and the relative contribution from the ``fission bottleneck" $x^\mathrm{f}_\mathrm{FS/FFP}$ increases at $A=120-130$.
The fission fragments produced above $N=82$ ($A\gtrsim 130$) spread out up to $N=126$  ($A\lesssim 195$).
For $A>160$, the fission contribution to the final production $x^\mathrm{f}_\mathrm{FFP}$ and the relative contribution of the ``fission bottleneck" $x^\mathrm{f}_\mathrm{FS/FFP}$ is higher with GEF than with SPY since the pre-freeze-out distribution is more asymmetric due to the more asymmetric FFD of \textsuperscript{246}Am.
At freeze-out, there is no more fission fragments produced by the ``fission bottleneck" and the ``fission roof".
The final abundance peak $A=130$ predicted by SPY model (Fig.~\ref{fig:allmYf}) is due to the high post-freeze-out mean fission yields around $A=130$ (Fig.~\ref{fig:FFD_SPYvsGEF_noNU}, orange curves).

$x^\mathrm{f}_\mathrm{FFP}$ is locally reduced around $A=130$ due to the accumulation of non-recycled material  along the neutron shell closure $N=82$.
This reduction is limited with SPY model since the pre- and post-freeze-out mean fission yields are higher than in the GEF case around $A=130$.
A low $x^\mathrm{f}_\mathrm{FFP}$ for $80<A<100$ could be expected since the mean fission yields are negligible for both FFD models.
However, in this case, there is no matter accumulation in this mass region, so that the low $80<A<100$ production ($Y^\mathrm{f}<10^{-4}$) occurring after freeze-out is essentially due to the asymmetric fission of seeds  located in the ``decay roof 0".
This leads to a non-negligible value of $x^\mathrm{f}_\mathrm{FFP}$ for $80<A<100$.
The drop of $x^\mathrm{f}_\mathrm{FFP}$ for $A>180$ is due to the accumulation of matter along the neutron shell closure $N=126$.

\subsubsection{Scenario~II}
\label{sec:FPsII}

In scenario~II, the reduced fission recycling leads to a non-negligible  $x^\mathrm{f}_\mathrm{FFP}$ that does not exceed 30\% only around $130<A<185$, irrespective of the FFDs model adopted (Fig.~\ref{fig:FFD_SPYvsGEF_NuEC}, red dotted curves).

For both FFDs models, the post-freeze-out mean fission yields is roughly similar than the one within scenario~I since the main contributors from the ``decay roof" regions are the same in both cases (see Fig.~\ref{fig:spygef_93104}) where $x_\mathrm{CN}$ reaches values up to 0.08 for \textsuperscript{282}No.
The fission seed contribution from  the ``fission roof" is small, so that the $A=180-190$ pre-freeze-out mean fission yields are negligible with the SPY model.
The GEF pre-freeze-out mean fission yields is roughly similar to the one found in scenario~I, though the $A=100-120$ peak is now found higher than the $A=160-180$ one.

The low neutron irradiation makes the contribution of the ``fission bottleneck" and the ``fission roof" relatively small within this scenario, leading to a total $x_\mathrm{FS}<0.25$ for 90\% the ejected mass (\textit{i.e.} $X_\mathrm{fiss,tot}^\mathrm{cum}<0.1$) (Fig.~\ref{fig:fisslim_Xcumfiss}).
The role played by the pre-freeze-out mean fission yields is consequently reduced.
It remains more difficult to overcome the neutron shell closure $N=82$ and more non-recycled material accumulates along this shell closure. For this reason,  the final abundance around $A=130$ is higher than in scenario~I with a negligible fission contribution to the $A\leqslant 130$ abundances.
As the SPY post-freeze-out mean fission yields is peaked around $A=120-160$, $x^\mathrm{f}_\mathrm{FFP}$ is maximal for $A=135-160$.
Compared to SPY, GEF pre- and post-freeze-out mean fission yields are higher for $A\gtrsim 160$  which leads to a higher $x^\mathrm{f}_\mathrm{FFP}$ in this mass region.

\section{Conclusion}
\label{sec:con}

In the present study, new improvements have been brought to our scission-point model, called SPY, to derive the fission fragment distribution for all  neutron-rich fissioning nuclei of relevance in  r-process calculations. 
These improvements include a phenomenological modification of the scission distance and an updated  smoothing procedure of the distribution to take into account effects that are neglected in the original model. 
Such corrections lead to a better agreement with experimental fission yields. 
In particular, SPY FFDs are now in overall good agreement with the experimental symmetric/asymmetric transition in the Th and Pa isotopic chains. 
Those yields at scission are also used to estimate the number of neutrons emitted by the excited fragments on the basis of different neutron evaporation models.
Our new fission yields significantly differ from to those predicted by the widely used GEF model.

The impact of fission on the r-process nucleosynthesis in neutron mergers is also re-analyzed. 
Two scenarios are considered, the first one with low initial electron fraction is subject to intense fission recycling, in contrast to the second one which includes weak interactions on free nucleons resulting in less pronounced fission recycling.
The efficiency of the r-process nucleosynthesis and the corresponding abundance distribution in a given mass element has been discussed essentially in terms of the initial neutron mass fraction $X_n^0$ which engulfs  information on both the initial electron fraction and entropy. 
The contribution fission processes may have to the final abundance distribution has been studied in detail in the light of newly defined quantitative indicators, namely indicators for the fission recycling ($X_\mathrm{fiss,tot}^\mathrm{cum}$), fission seeds ($x_\mathrm{FS}$) and fission fragment progenitors ($x^\mathrm{f}_\mathrm{FFP}$). 
The various regions of the nuclear chart responsible for fission recycling during the neutron irradiation as well as after freeze-out have been identified for the nuclear physics adopted in the present calculations, what we defined as the ``decay roofs", the ``fission bottleneck" and the ``fission roof". 
Within the ``fission bottleneck'' region, Am isotopes around $A=285$ play a key role by limiting the nuclear flow to higher elements and strongly contributing to the fission recycling, mainly due to the low fission barriers predicted by the BSk14 interaction in this region. 
Above $Z=110$, the so-called ``fission roof'' prohibits the production of super-heavy elements. 
``Decay roofs'' essentially recycle heavy material after the neutron irradiation freezes out when the material accumulated at the $N=184$ shell closure decays back to the valley of $\beta$-stability.

When considering an efficient r-process (scenario I), fission fragments are found to contribute almost entirely to the final abundances of nuclei with $100 \lsimeq A \lsimeq 180$, while in scenario~II where weak interactions on nucleons initially limit the number of free neutrons, such a contribution does not reach more than 30\% and is concentrated in the $A=140-180$ region depending on the FFD model adopted.
Calculations based on the SPY and GEF FFDs have been compared for both r-process scenarios.

The present study allows us to have a deep insight on the role of fission processes in hydrodynamical r-process simulations on the basis of sound well-defined quantitative indicators. 
Fission remains an important nuclear process taking place in binary NS mergers, though major nuclear as well as astrophysical uncertainties still need to be further addressed before concluding on the definite role of fission processes. 
These concern in particular the determination of the fission probabilities for all the nuclei potentially involved during the r-process as well as a detailed description of neutrino absorption and hydrodynamics in simulations. 
Much progress remains to be done along these two directions. 

\section*{Acknowledgments}

J.-F.L. and S.G. acknowledge financial support from FNRS (Belgium). 
This work was supported by the Fonds de la Recherche Scientifique (F.R.S.-FNRS) and the Fonds Wetenschappelijk Onderzoek - Vlaanderen (FWO) under the EOS Project nr O022818F. 
The present research benefited from computational resources made available on the Tier-1 supercomputer of the F\'ed\'eration Wallonie-Bruxelles, infrastructure funded by the Walloon Region under the grant agreement nr 1117545.
H.-T.J. acknowledges financial support from Deutsche Forschungsgemeinschaft  (DFG, German Research Foundation) through Sonderforschungsbereich (Collaborative  Research  Center) SFB-1258 ``Neutrinos and Dark Matter in Astro- and Particle Physics (NDM)'' and under Germany’s Excellence Strategy through Cluster of  Excellence ORIGINS (EXC-2094)—390783311, and from the European Research Council through Grant ERC-AdG No. 341157-COCO2CASA.
A.B. acknowledges support by the European Research Council (ERC) under the European Union's Horizon 2020 research and innovation programme under grant agreement No. 759253, and by Deutsche Forschungsgemeinschaft (DFG, German Research Foundation) - Project-ID 279384907 - SFB 1245 and - Project-ID 138713538 - SFB 881 (``The Milky Way System'', subproject A10).

\bibliography{biblio}

\end{document}